\documentclass[fleqn,12pt]{olplainarticle}

\usepackage{float}

\newcommand{\Graph}{safe progression graph}
\newcommand{\mG}{\mathcal{G}}

\title{Controlling ball progression in soccer}

\usepackage{amsthm}
\theoremstyle{definition}
\newtheorem{definition}{Definition}[section]
\newtheorem{remark}{Remark}

\newtheorem{ex}{Example}

\author[1]{Catherine Pfaff}
\author[1]{Emily Hunter}
\author[1]{Haozhi Hong}
\author[1]{Daniel Forestell}
\author[1]{Ari Fialkov}
\author[1]{Zoey Drassinower}
\author[2]{Timothy Chan}

\affil[1]{Queen's University}
\affil[2]{University of Toronto}


\begin{abstract}
\indent In this paper, we examine how soccer players can use their spatial relationships to control parts of the field and safely move play up the field via chains of ``safe configurations,'' i.e. configurations of players on a team ensuring the possessor of the ball has a collection of open passing options all connected by open passing lanes. An underlying philosophy behind our work is that it is most difficult to disrupt an attacking team's progression forward (with the ball) when this attacking team has multiple ``good'' options of how to proceed at each moment in time. We provide some evidence of this. Our main construction is a directed weighted graph where the nodes encode the configurations of players, the directed edges encode transformations between these configurations, and the weights encode the relative frequencies of the transformations. We conclude with a few applications and proposed further investigations. We believe that our work can serve as a launching platform for significant further investigation into how teams can ``safely progress'' the ball up the field, strategy development, and sophisticated decision making metrics. For coaches and players, we aim to streamline the process of moving safely up the field. In particular, we aim to construct a framework for creating new successful patterns of movement and aim that our framework allows for the development of new strategy. At the same time, our work could help identify configurations on the field which often result in turnovers or that allow for a lot of strategic flexibility (also impeding defensive containment of the attacking team).

\bigskip

$^\ast$ Because the contributions of the female authors to this paper were by no means less than those of the male authors, we have chosen to reverse convention and to list the authors in reverse alphabetical order to ensure that the female authors were not listed only after the male authors.
\end{abstract}

\begin{document}

\flushbottom
\begin{titlepage}
\maketitle
\thispagestyle{empty}
\end{titlepage}

\section{Introduction}

In this paper, we examine how soccer players can use their spatial relationships to control parts of the field and safely move play up the field via chains of ``safe configurations,'' i.e. configurations of players on a team ensuring the possessor of the ball has a collection of open passing options all connected by open passing lanes. An underlying philosophy behind our work is that it is most difficult to disrupt an attacking team's progression forward (with the ball) when this attacking team has multiple ``good'' options of how to proceed at each moment in time. We provide some evidence of this in \S \ref{sec:future_possession_probability}.

One of the largest early advances in soccer analytics was that of Sarah Rudd \cite{rudd}, where she used chains of events (such as dribbles or passes) within a soccer game to place values on individual actions. Van Roy, Yang, De Raedt, and Davis \cite{van2021analyzing} built upon this work and incorporated it into defensive strategy. One of the most sophisticated notions building upon Rudd's notion of an ``expected threat'' was the ``expected possession value'' developed by Fernandez, Bornn, and Cervone \cite{epv_2019, epv_2020}. There are also the g+ \cite{gPlus} and VAEP \cite{decroos2019actions} models.

With the arrival of tracking (spatial-temporal) data and applications of physics principles, William Spearman \cite{pitchcontrol} introduced ``pitch control,'' a model quantifying the probability that a given team will have possession of the ball if it were at a location $x$.

More recent papers have looked at team motion, including that \cite{goes2021tactics} studies the motion of a team as a whole, with a focus on team synchronization. Along a slightly different tack, \cite{beernaerts_baets_lenoir_weghe_2020} focused on spatial movement pattern detection.

In this paper we build off of the work of Rudd, Spearman, and others to tackle a new question, namely how the configurations of players on a field (and their evolution over time) can enable a team to ``safely'' progress the ball from the defensive third to the attacking third. To our knowledge, our work is unique in studying the evolution of configurations of players on an attacking team as they move from the defensive third to attacking third, particularly with the flexibility our framework provides for developing new strategies.

As a first step toward defining what we will call ``safe configurations,'' (Definition \ref{def:safe_configuration_players}) we build on Spearman's notion of pitch control, as defined in \cite{pass_probabilities, pitchcontrol}, to define ``passable regions'' where a particular team is most likely to control the ball upon arrival (please see \S \ref{sec:passable_regions}). For a configuration of players on an attacking team to be considered ``safe,'' we will require that a suitable combination of passing lanes (please see \S \ref{sec:identifying_passing_cliques}) is within the attacking team's passable region.

We next define in \S \ref{sec:safe_progression_graph} a directed weighted graph, which we will call the ``safe progression graph,'' where a node will represent a safe configuration, a directed edge between two nodes will represent a transformation between the configurations represented by the respective nodes, and weights indicate a transformation's frequency.

As mentioned before, an underlying philosophy behind our work, that we provide some evidence of in \S \ref{sec:future_possession_probability}, is that it is most difficult to disrupt an attacking team's progression forward (with the ball) when this attacking team has multiple ``good'' options of how to proceed at each moment in time.
This manifests itself in our work as follows:  In \S \ref{sec:safe_progression_graph} we define safe configuration nodes by requiring that the player with the ball is part of a complete graph whose vertices are players on their team and whose edges are passing lanes between these players that are ``safe'' in that they lie entirely within the attacking team's region of control. This in fact encodes 2 steps of options. In the first step the player with the ball can safely pass to each other player representing a vertex in the complete graph. In the second step the player who receives the ball can again pass to any other player representing a vertex in the complete graph.

We believe our work can serve as a launching platform for significant further investigation into how teams can ``safely progress'' the ball up the field, strategy development, and sophisticated decision making metrics. For coaches and players, we aim to streamline the process of moving safely up the field. In particular, we aim to construct a framework for creating new successful patterns of movement and aim that our framework allows for the development of new strategy. At the same time, our work could help identify configurations on the field which often result in turnovers or that allow for a lot of strategic flexibility (also impeding defensive containment). Finally, we aim that, through analysis of our structures, one can help improve players' awareness of their position relative to others and give a context to apply this awareness to. In \S \ref{sec:conclusion_and_further_questions} we discuss some potential further investigations and coaching applications of our work presented here.

\section{Data}\label{sec:data}

\begin{figure}[ht]
    \centering
    \begin{subfigure}[b]{0.45\textwidth}
        \centering
        \includegraphics[width=\textwidth]{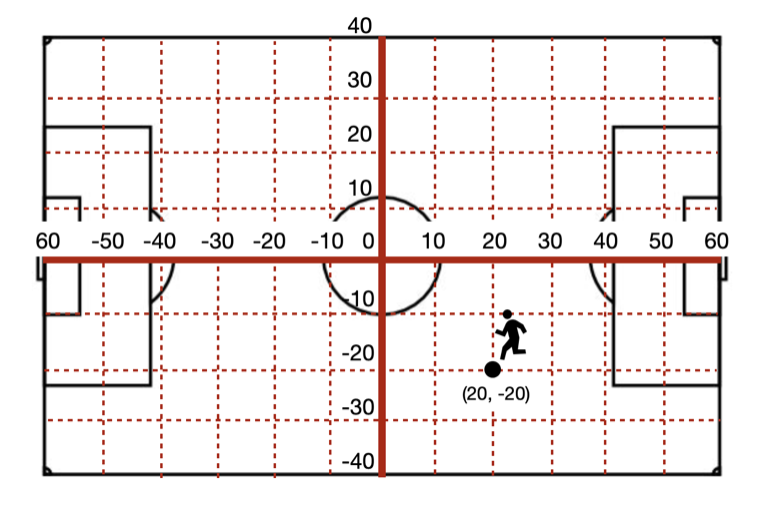}
        \caption{}
        \label{fig:field_grid_1}
    \end{subfigure}
    \begin{subfigure}[b]{0.455\textwidth}
        \centering
        \includegraphics[width=\textwidth]{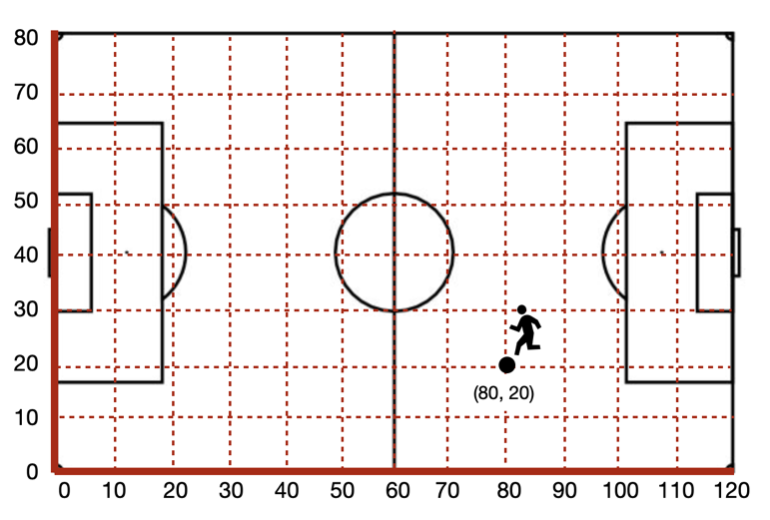}
        \caption{}
        \label{fig:field_grid_2}
    \end{subfigure}
    \caption{(a) The coordinate system found in the tracking data. (b) The coordinate system used for player and ball positions after applying the transformation.}
    \label{fig:field_grid}
\end{figure}

The data used in this paper is Signality-provided tracking data for games from the men's Allsvenskan 2021 season. Tracking data for each game is organized into rows, where each row is a \emph{frame}. There are 25 frames of data captured per second. We used the following data elements from each frame in our analysis:
\begin{itemize}
    \item Whether the frame belongs to the first half or second half of the game.
    \item The frame index, which depends on the match time, in milliseconds.
    \item Data for each of the players on the home and away team. This includes their jersey number, $(x, y)$ field position,
    and their current speed.  (Please see Figure \ref{fig:field_grid_1}.)
    \item Information about the ball, including its $(x, y)$ coordinate and which team is in possession (please again see Figure \ref{fig:field_grid_1}).
\end{itemize}

Our analysis required each player's velocity in both the $x$ and $y$ directions for each frame. This data was derived using their speed and the change in their position from the previous frame. It is important to note at this point that there are large gaps in the data (primarily pertaining to the location of the ball). While the missing ball location is unlikely to significantly impact the player velocity computations, frames missing player locations did occur and could impact velocity computations.

The original tracking data stored each player's $(x, y)$ coordinates according to a graph where the origin is in the center of the field, as in Figure \ref{fig:field_grid_1}. To make it consistent with other sources of tracking data, we converted it to the $120 \times 80$ grid shown in Figure \ref{fig:field_grid_2} via a transformation, such as $(x,y)\mapsto (x+60,y+40)$.\footnote{There are small variations in the pitch size impacting the transformation. For example, the most typical in the data we used was $105 \times 68$. It is noteworthy that the field size (and coordinate system) after the transformation is always the same.}

\section{The Safe Progression Graph $\mG_{\text{safe}}$} \label{sec:safe_progression_graph}

In this section we define the safe progression graph $\mG_{\text{safe}}$. The process is as follows:
\begin{enumerate}
    \item We define in \S \ref{sec:passable_regions} ``passable regions'' (a modification of Spearman's pitch control model).
    \item In \S \ref{sec:identifying_passing_cliques} we define a ``safe configuration'' of players to be one with a ``passing clique,'' i.e. the player in possession of the ball is part of a complete graph where the vertices represent teammates and the edges represent passing lanes contained entirely within the team's passable region.
    \item From the safe configurations we define in \S  \ref{sec:Gaussian_Blurred_passing_matrix} ``passing matrices'', on which we will perform a clustering algorithm to form the nodes of $\mG_{\text{safe}}$.
    \item The distance metric and matrix, as well as other aspects for the clustering, are defined in \S  \ref{sss:clustering_cliques}
    \item In \S \ref{sec:safe_progression_graph} we finally define the safe progression graph $\mG_{\text{safe}}$ itself.
\end{enumerate}

\subsection{Passable Regions} \label{sec:passable_regions}

In order to define safe passing lanes and cliques, to then define safe configurations, we will need a model dividing the soccer pitch into probabilistic zones of control for each team. For this purpose we define the passable regions model, which is similar to Spearman's pitch control model in that it assigns points on the field to certain teams by determining which team has a player who could be first to reach that location. This takes into account the players' current velocity, direction, and maximum acceleration (please see \S \ref{ss:computations}). Unlike the pitch control model, which assumes that the ball is at a particular location $(x, y)$ and then assigns that location to a particular team, the passable regions model also calculates the time required for the ball to reach $(x, y)$ from its current location. This will be important for our purposes because we focus on situations with ``safe'' passing lanes. To illustrate the significance, one can imagine a situation where it looks like there is a clear passing lane from player A to player B and player B controls a region for ``receiving'' the ball, but the ball is intercepted because an opposition player could reach that location of the passing lane before the ball could, so the region player B controls is only useful for aerial passes. We note that this kind of situation is taken into consideration by Spearman et al in \cite{pass_probabilities}, but was not included in the application of the Pitch Control Function \cite{pass_probabilities, pitchcontrol}. See Definition \ref{def:passable region} for the full definition of a team's passable region.

\subsubsection{Computations}\label{ss:computations}

We computed the times for all players and the ball to reach a particular location $(x, y)$ using an initial velocity and constant deceleration for the ball, as well as estimates of acceleration and the maximum velocity for a player. As we found it did no have a significant enough impact on our computations, we left out consideration of the time it takes a player to change direction (turn). We could then determine whether it would be possible for a player to reach this location $(x, y)$ before the ball. When the model is applied to a moment in the game we obtain images as seen in Figure \ref{fig:passable_regions}.

Since we found little supporting research for the physics parameters involved in passing the ball, we determined the physics parameters as follows (using the data of Mjällby AIF's games as the sample set).

To obtain the ball's initial speed and deceleration, using the start and end time of a pass given in the event data, we:

\noindent $\bullet$ estimated the ball speed by the central difference method, and then

\noindent $\bullet$  changed the start time of each pass to where it reached its maximal speed, and then

\noindent $\bullet$  determined a ``good'' subset of the passing data by imposing the following 3 condition on passing data:

1. we required that the data of each pass (i.e. the data in the interval from the start time to the end time) contains $\geq 20$ frames (0.8s) and

2. required that the data satisfies smoothness, i.e. that the speed difference between any 2 adjacent frames is always $<2$ m/s and

3. required that the average speed during the 1st half of the data during a pass is larger than during the 2nd half.

The initial ball speed is then taken to be the average of the set of maximal speeds of the sets of ``good'' passing data (there is at most 1 ``good set'' for each pass and then 1 maximal speed for each ``good set'', we average these maximal speeds). For the deceleration, we use linear regression on the ``good'' passing data and take the slope of a fit as the deceleration of a pass. The ball deceleration is taken as the average deceleration of all of the ``good'' passes.
Through this methodology we obtain an initial ball speed of 16.20 m/s and ball deceleration of 7.93 m/$\text{s}^2$.

We now explain how we found the maximal speed and maximal acceleration of a player that we use in our computations. We use the values of a maximal speed of 10 m/$s^2$ and a maximal acceleration of 4.98 m/s (taken from \cite{Loturco2019}), as they are close to those we determined using the following methodology. We use the central difference method to estimate the acceleration of a player (the speed of a player is taken from the data).
The speed and acceleration in all Mjällby AIF's games is plotted and we obtain that the
top 1$\%$ of the accelerations are above 4.779 m/$s^2$, and
the top 0.5$\%$ of the accelerations are above 5.718 m/$s^2$ The maximal speed is 11.13 m/s and
the top 1$\%$ of this speed is above 7.12 m/s.

Finally, the horizontal distance covered by the ball while in contact with the foot is approximately 2/3 the diameter of the ball \cite{Asai2002}. As the average diameter of a soccer ball is $22cm$, this means that the ball travels $14.67cm$ while in contact with the foot.

Supposing the ball must travel a total distance of $d$, we set $d_r = d - 0.1467$m as the remaining distance to be travelled after the ball leaves the foot. The time $t_b$ taken for the ball to travel the remaining distance $d_r$ is then calculated via the 1-dimensional calculus computation given below  with the given initial velocity $v_b$ and deceleration $-a_b$ computed above. The time is set to infinity if the ball's velocity reduces down to zero before travelling the distance of $d_r$. The time is set to 0 if $d<1$m.

We now give the calculus computations we used. We let   $d_{remaining} = max(d-0.1467, \; 0)$ and then define
\begin{equation}
    \Delta :=  v_{b}^2 + 2a_{b}d_{remaining}
\end{equation}

The time required for the ball to reach the given location was then:
\begin{equation}
    t_{b} =
    \begin{cases}
        0 & d<1 \\
        \frac{\sqrt{\Delta}-v_b}{a_b}  & \Delta \geq 0 \text{ and } d\geq 1\\
        \infty & \text{otherwise}
    \end{cases}
\end{equation}

Next we calculate the time required for a player to reach a particular location on the field. We assume that the player turns around and then moves toward the final location $\vec{r}$. The initial velocity $\vec{v_0}$ of a player can be taken from the the tracking data. We set the velocity toward the final location to be
\begin{equation*}
    v_{toward} = ||\vec{v_0}||\cos{\theta}
\end{equation*}
where $\theta$ is the angle that the player needs to turn. The time $t_p$ taken for the player to travel is then calculated via the 1-dimensional calculus computation of the time it takes a player to reach a given distance $||\vec{r}||$ with initial velocity $v_{toward}$ and then the maximal speed $\tilde{v}_p$ and maximal acceleration $\tilde{a}_p$ computed above. We set the time to be 0 if $||\vec{r}||<1$.

We defined $v_{max}$ as
\begin{equation}
    v_{max} = max(v_{toward}, \; \tilde{v}_p)
\end{equation}
\noindent and then set
\begin{equation}
    d_{max} = \frac{(v_{max})^2-(v_{toward})^2}{2\tilde{a}_p}.
\end{equation}
The time required for the player to reach the given location is then
\begin{equation}
    t_{p} =
    \begin{cases}
        0 & ||\vec{r}||<1 \\
        \frac{\sqrt{(v_{toward})^2 + 2\tilde{a}_p||\vec{r}||}-v_{toward}}{\tilde{a}_p}  & 1 \leq ||\vec{r}|| <  d_{max}\\
        \frac{v_{max}- v_{toward}}{\tilde{a}_p} + \frac{||\vec{r}||-d_{max}}{v_{max}} & \text{otherwise}
    \end{cases}
\end{equation}

\subsubsection{The passable region definition}\label{ss:passable_region_definition}

\begin{figure}[ht!]
    \centering
    \begin{subfigure}[b]{0.8\textwidth}
        \centering
        \includegraphics[width=80mm]{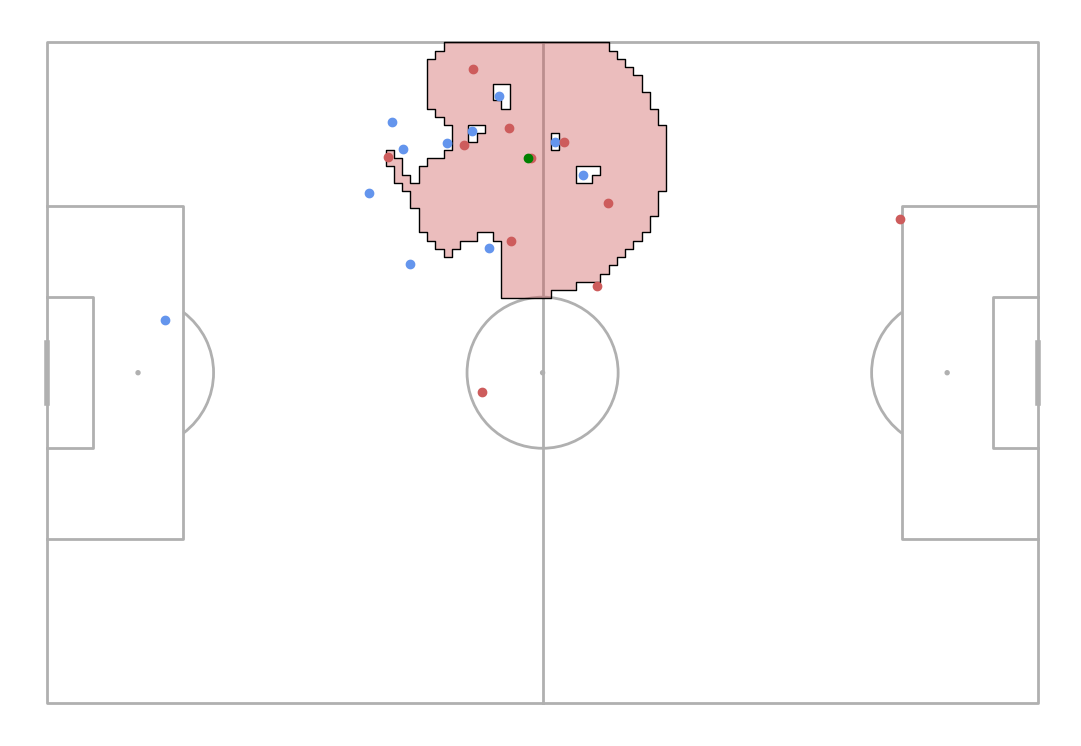}
    \end{subfigure}
    \caption{This figure illustrates the passable region (shaded in red) for the red team at a given moment in time in a game. The two teams are shown as dots in red and blue, and the ball is green.}
    \label{fig:passable_regions}
\end{figure}

Every 1x1 metre square on the field has been assigned a colour, namely red if the attacking team can reach that square first or the ball can reach there first (and actually reach that location) and white otherwise. Using these equations, and the resulting coloured squares, we define a ``passable region'' as follows.

\begin{definition}[Passable region]\label{def:passable region}
The (representation of the) \emph{passable region} for a given team at a given moment in a game is the union of the following sets of squares (where squares refer to the squares of Figure \ref{fig:passable_regions} representing 1x1 metre areas of the pitch):
\begin{itemize}
    \setlength{\itemindent}{5em}
    \item the set of squares that a player from that team can reach before any other player and
    \item the set of squares that the ball reaches before any player.
\end{itemize}
In terms of the visual representation, this is the union of all red squares.

\end{definition}

In short, the passable region for a team represents the area of the field where that team can play the ball and be reasonably confident of maintaining possession.

\begin{remark}
Our passable regions model additionally gives a refinement of the space controlled by a team. For instance, if the ball is in one of the corners, it would not matter much which team ``controlled'' the opposite corner according to the model, as a player would not be able to reach there before the ball. If we had counted this ``unreachable'' space, it might falsely inflate the space a team is said to control.
\end{remark}

\subsection{Identifying passing cliques \& safe configurations of players} \label{sec:identifying_passing_cliques}

Next we define in Definition \ref{def:safe_passing_lane} a line segment connecting the locations of two players on the same team to be a ``safe passing lane'' when it only passes through that team's passable region. Safe passing lanes represent passes which have a high probability of success. We did not include passes into empty space for a player to run into or aerial passes, although a more advanced model could incorporate this.

\begin{definition}[Safe passing lane]\label{def:safe_passing_lane}
Consider locations of players at a given point in time to be as in the coordinate system of Figure \ref{fig:field_grid_2}. Suppose that $p_1$ and $p_2$ are player locations at a given point in time for two players on the same team. We call the line segment between $p_1$ and $p_2$ a \emph{safe passing lane} (between $p_1$ and $p_2$) if it is contained entirely within the passable region of the team containing the players represented by $p_1$ and $p_2$.
\end{definition}

\begin{figure}[ht!]
\centering
    \begin{subfigure}[b]{0.45\textwidth}
        \centering
        \includegraphics[width=\textwidth]{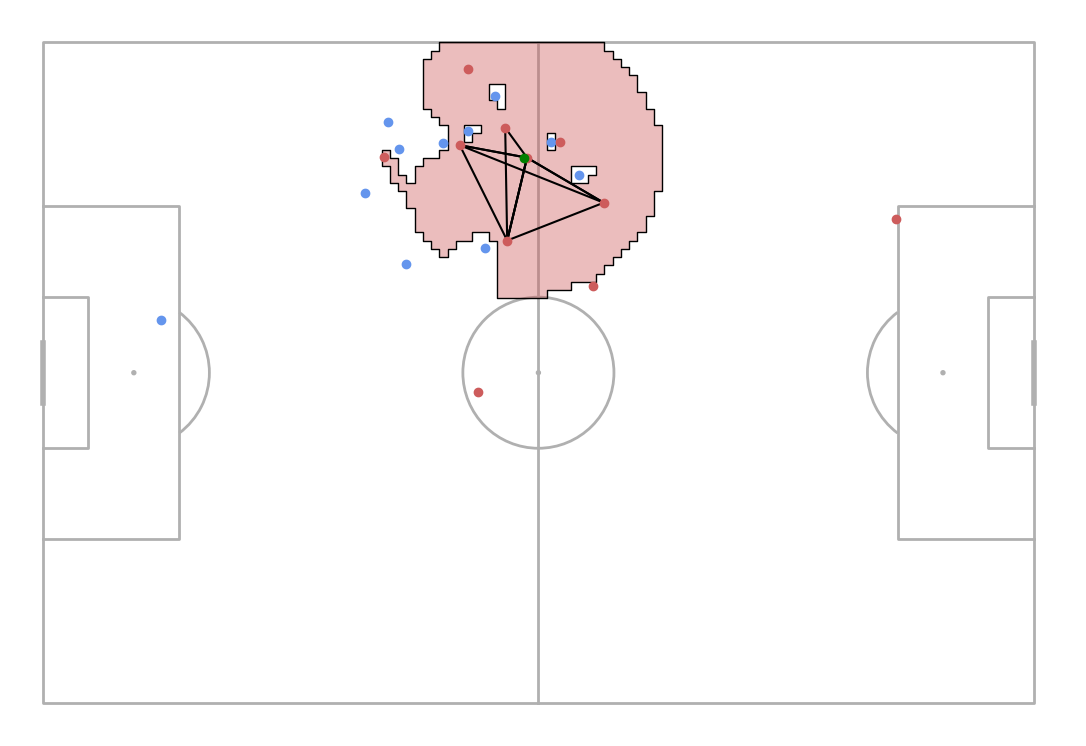}
        \caption{}
        \label{fig:passing1}
    \end{subfigure}
    \begin{subfigure}[b]{0.45\textwidth}
        \centering
        \includegraphics[width=\textwidth]{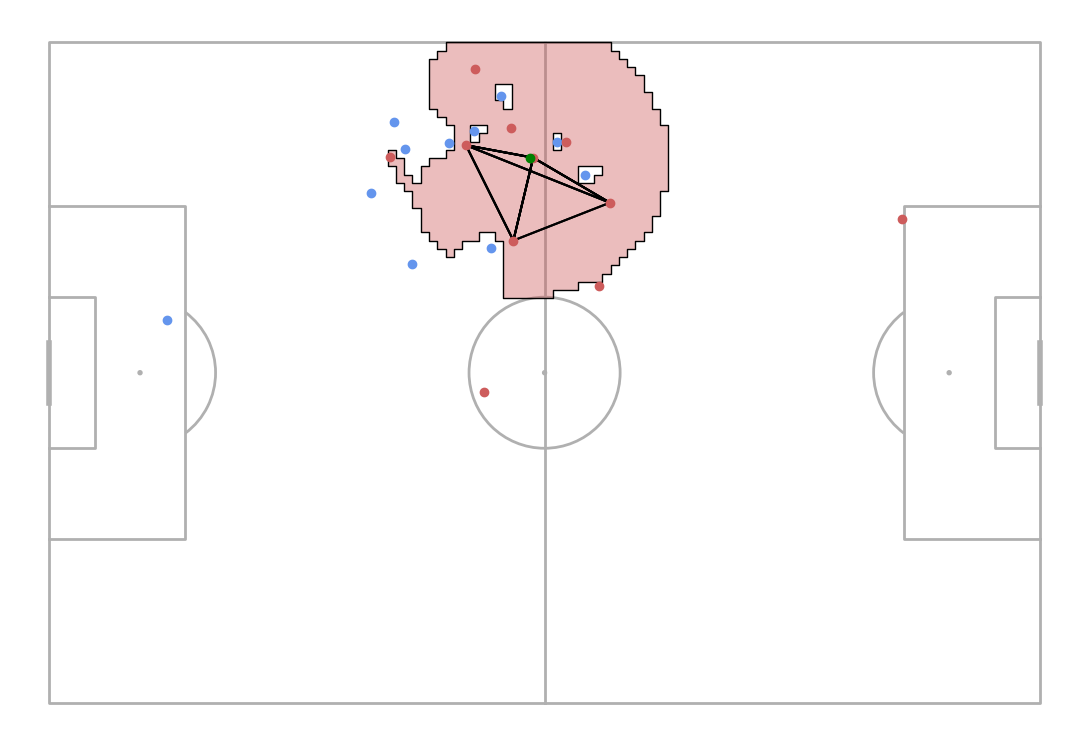}
        \caption{}
        \label{fig:passing2}
    \end{subfigure}
    \hfill
    \caption{(a) Using the passable regions model, we defined ``safe passing lanes'' between players and then ``passing cliques.'' The left-hand image depicts all passing cliques for the red team. Note, there is a vertex for each player of the red team that is part of a passing clique and these vertices are connected by an edge precisely when that edge only runs through the red team's passable region. The locations of the blue players, and those red team players not in passing cliques, are not represented by vertices but do impact the red team's passable region, and thus the possible ``safe passing lanes.'' (b) Here we include only a single maximal passing clique. The configuration of players (and play) depicted in this figure had 2 maximal passing cliques.}
    \label{fig:passing_cycles}
\end{figure}

For each instance of time we define a graph, called the ``passing graph,'' and from this ``safe configurations,'' according to Definition \ref{def:passing_graph}. (Please see Figure \ref{fig:passing_cycles} for accompanying images.)

\begin{definition}[Passing graph, passing clique, safe configuration of players]\label{def:passing_graph} \label{def:safe_configuration_players}
We are again considering locations of players to be as in the coordinate system of Figure \ref{fig:field_grid_2}. Given a team and their passable region at a moment in time, we define a \emph{passing graph} as the graph which has:

$\bullet$ a vertex for the location of each player on that team and

$\bullet$ an edge between 2 vertices if the line segment between the players represented by those 2 vertices is a safe passing lane, in the sense of Definition \ref{def:safe_passing_lane}.

\noindent By a \emph{passing clique} we will mean a subgraph of the passing graph that

$\bullet$ has $\geq 3$ vertices,

$\bullet$ is a complete graph in a graph-theoretic sense, and

$\bullet$ satisfies that one of the vertices represents a player in possession of the ball (i.e. the ball is within 1m of this player).

We say that a collection of players on a team is in a \emph{safe configuration} if one of these players is in possession of the ball and that player in possession of the ball is part of a passing clique formed in conjunction with other players from that collection.

\end{definition}

\begin{remark}[Passing clique interpretations]
We mentioned in the introduction how having a complete graph leads to multiple steps of multiple options, making it hard for the opposition to halt progression. We mention another interpretation now. When a graph component is incomplete rather than complete it implies an opposing player is standing within the graph component or can block a number of the passes between the players.
It is for these reasons that we restrict our attention to passing cliques.
\end{remark}

\subsection{Clustering safe configurations of players} \label{sec:clustering_cliques}

The goal now will be to identify and categorize configurations of players leading to passing cliques arising in games.

\subsubsection{The (Gaussian-blurred) passing matrix} \label{sec:Gaussian_Blurred_passing_matrix}

As a first step in the identification and categorization of player configurations, we transform each data point into a matrix that encodes the proximity of players, the location of the ball, and the passing cliques. For each safe configuration of players, we then define a $120 \times 80$ ``passing matrix'' representing the players and their spatial relationships. The values for the entries were chosen using an ``eye test.''

\begin{definition}[Passing matrix]\label{def:passing_matrix}
Given a passing graph, the \emph{passing matrix} $[x_{ij}]$ is a $120 \times 80$ matrix defined as follows. The $i,j$ entry $x_{ij}$ corresponds to the square $(i, j), (i+1, j), (i, j+1), (i+1, j+1)$ on the field and is:

\noindent $\bullet$ 0 if there is no attacking team player of a passing clique nor the ball in the corresponding square.

\noindent $\bullet$ 5 if either there is an attacking team player within a clique but without the ball or the ball without an attacking team player.

\noindent $\bullet$ 10 if the square contains the attacking team player with the ball and this player is part of a passing clique.

\end{definition}

\begin{remark}[Passing matrix entries]
Note that entries of the passing matrix associate high values to areas of the field with players of passing cliques. Also, it is possible that the player possessing the ball is not in the same grid square as the ball (as possession is computed using a radius of 1 instead of a same-square requirement). Consequently, the ball and possessor of the ball may add weight to different matrix entries.
\end{remark}

If one clustered the vectors determined by the passing matrices information would be lost as to how close squares were to each other. To add this information we apply a Gaussian filter to the matrix.
This is achieved by the process described in Definition \ref{def:gaussian_passing_matrix}.

\begin{definition}[Gaussian-blurred passing matrix]\label{def:gaussian_passing_matrix}
Suppose $P=[x_{ij}]$ is a passing matrix. We define a 2-variable function $g: [0,  120] \times [0,  80] \rightarrow \mathbb{R}$, with elements of the domain representing field locations (see Figure \ref{fig:field_grid_2}), as follows. Let $n_{ij}$ denote the Gaussian distribution with mean at $(i, j)$, covariance $\big(\begin{smallmatrix}
  2 & 0\\
  0 & 2
\end{smallmatrix}\big)$, and weight $x_{ij}$. We define $g$ by $g(a,b)=\sum_{ij} n_{ij}(a,b)$ for each $(a,b)$ in $[0,  120] \times [0,  80]$, where the $i,j$ range over integers satisfying $0\leq i\leq 199$ and $0\leq j\leq 79$. Then the \emph{Gaussian-blurred passing matrix} $G(P)=[a_{ij}]$ is defined by $a_{ij}=g(i, j)$ for each $i \in [0, 199], j\in[0, 79]$. $G(P)$ will be a $120 \times 80$ matrix.

\end{definition}

\begin{figure}[ht!]
     \centering
     \begin{subfigure}[b]{0.32\textwidth}
         \centering
         \includegraphics[width=\textwidth]{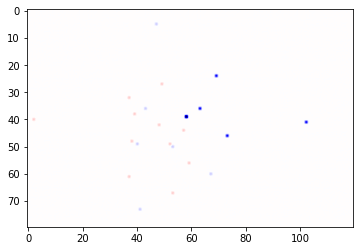}
         \caption{}
         \label{fig:step1}
     \end{subfigure}
     \hfill
     \begin{subfigure}[b]{0.32\textwidth}
         \centering
         \includegraphics[width=\textwidth]{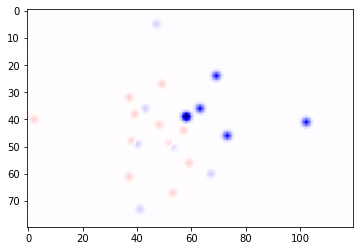}
         \caption{}
         \label{fig:step2}
     \end{subfigure}
     \hfill
     \begin{subfigure}[b]{0.32\textwidth}
         \centering
         \includegraphics[width=\textwidth]{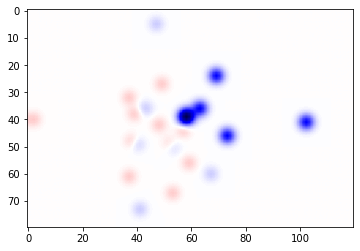}
         \caption{}
         \label{fig:step3}
     \end{subfigure}
     \caption{(a) Information about a passing clique is encoded in a passing matrix, as in Definition \ref{def:passing_matrix}. For simplicity, these images represent the Gaussian blurred passing matrix for a single maximal passing clique. Red players are the defensive team and blue players are the attacking team. The blue players of the maximal passing clique are slightly darker as they have larger entries in the matrix. The player with the ball is the darkest. These images are visual representations of $80\times 120$ matrix where the colours show the relative values of the matrix entries. (b) This image represents a Gaussian-blurred passing matrix, but with the covariance matrix of $\big(\begin{smallmatrix}
        1 & 0\\
        0 & 1
    \end{smallmatrix}\big)$ used. In part a), there were exactly 22 coloured cells in the matrix. Now, the cells around the players have been coloured as well according to their proximity to different players as a result of Gaussian blurring. For example, at roughly $(60, 40)$ there is a blue player with the ball. This player has the largest associated matrix entry and so has the darkest and biggest associated dot. (c) This image again shows a Gaussian-blurred passing matrix, but using a covariance matrix of
    $\big(\begin{smallmatrix}
        2 & 0\\
        0 & 2
    \end{smallmatrix}\big)$.}
    \label{fig:matrices_3_steps}
    \end{figure}

An example Gaussian-blurred passing matrix is shown in Figure \ref{fig:matrices_3_steps} c, where a colour scale is used to show the matrix values. The decision to maximize the passing matrix entry that had the ball results in all entries around the ball increasing, as seen in Figure \ref{fig:matrices_3_steps} c.

This function, $G$, was implemented in Python using the \texttt{transforms.GaussianBlur()} function from the \texttt{pytorch} library \cite{NeurIPS_pytorch}.
For more information about the Gaussian filter, one can reference \cite{gaussian_blur2}. The \texttt{skimage} Python library also implements a Gaussian filter which is described in \cite{gaussian_blur}.

\subsubsection{Clustering the safe configurations} \label{sss:clustering_cliques}

Before clustering the configurations, to reduce the computational resources needed for clustering, we partition the frames containing safe configurations into 8 sets. Given a frame $X$, let $n(X)$ denote the total number of players that are involved in at least one of the passing cliques in that frame. We remove from consideration all frames $X$ for which $n(X)>6$. We then partition the set of all remaining frames using $n(X)$ and whether or not the goal keeper is contained in a passing clique. Since a passing clique must contain $\geq 3$ players, this gives 2 partition elements for each value in $\{3,4,5,6\}$, namely one for when the goalie is included in a clique and one for when the goalie is not included in a clique. We then apply the clustering algorithm separately to each partition element, obtain a list of clusters for each partition element, and combine each of these lists of clusters into a single list.

To implement a clustering algorithm, we must first define a notion of the distance between two Gaussian-blurred passing matrices. This distance captures how similar two player configurations (plays) are in terms of both player location and the location and orientation of the passing clique. Since there is a relationship between passing cliques and (Gaussian-blurred) passing matrices, this distance between the matrices can also be used to describe the similarity between passing cliques.

In keeping with the terminology of clustering, each (Gaussian-blurred) passing matrix will be referred to as a ``data point" for the remainder of this section.

\begin{definition}[Passing matrix distance $d_P$, cluster diameter]\label{def:passing_clique_distance}
Given two data points, $P_1$ from frame $F_1$ in the tracking data and $P_2$ from frame $F_2$, we define the \emph{passing matrix distance} between $P_1$ and $P_2$, denoted $d_P(P_1, P_2)$, as the $L^1$ distance between the Gaussian-blurred passing matrices $G(P_1)$ and $G(P_2)$.

By the \emph{diameter} of a set of data points (such as those in a safe configuration cluster), we will mean the maximum passing matrix distance between any 2 data points in the set.
\end{definition}

\begin{remark}[Alternatives to the Gaussian blur]
While the $L^1$ metric on its own does not encode information on whether two cells represent adjacent squares in the field, we artificially insert this information using a Gaussian blur, as previously discussed. One possibility that we have not yet explored is to view player locations in the field as spikes in a probability distribution and then to take the Wasserstein distance between two probability distributions. We chose to proceed with the Gaussian blur because it was already implemented in python and has low computational complexity.
\end{remark}

These passing matrix distances become entries in a new, very large, distance matrix according to Definition \ref{def:distance_matrix}.

\begin{definition}[Distance matrix]\label{def:distance_matrix}
Suppose the set of tracking data creates $n$ Gaussian-blurred passing matrices with a fixed order. Then the \emph{distance matrix} is an $n \times n$ upper triangular matrix where the $(i, j)$ entry is the passing matrix distance between the $i$th and $j$th Gaussian-blurred passing matrices. The passing matrix distance, $d_P$, is a symmetric relation, and so
the distance matrix is reduced to be upper triangular.
\end{definition}

\begin{definition}[Safe configuration cluster]\label{def:cluster}
Given a set of tracking data on which we have applied a suitable clustering algorithm,
we will refer to the clusters returned from the algorithm as the \emph{safe configuration clusters}.

\end{definition}

\subsection{The safe progression graph and its adjacency matrix}\label{sec:safe_progression_graph}

In the previous section we defined a ``safe configuration cluster.'' In this section we define the ``safe progression graph'' $\mG_{\text{safe}}$, which encodes how frequently teams have historically moved between which clusters, i.e. shifted between the configurations within 2 (possibly different) clusters.

\subsubsection{The safe progression graph $\mG_{\text{safe}}$}\label{sss:safe_progression_graph}

\begin{definition}[The safe progression graph $\mG_{\text{safe}}$]\label{def:safe_progression_graph}
Given a set of tracking data and parameters defining the passing matrix distance, the \emph{safe progression graph $\mG_{\text{safe}}$} will have:

 $\bullet$ A \emph{node} for each safe configuration cluster.

 $\bullet$ A \emph{set of directed edges $E(A, B)$} defined by there being a directed edge from Node $A$ to Node $B$ if the team in question has ever in the data set within 10 seconds transformed from the safe configuration cluster represented by $A$ to the safe configuration cluster represented by $B$.

 $\bullet$ A \emph{set of weights $w(A, B)$}, one for each edge defined as follows. The weight $w(A, B)$ of a safe progression graph edge $E(A, B)$ is the number $x \in (0, 1]$ that is the proportion of the transformations from the safe configuration cluster of Node $A$ which go to the safe configuration cluster of node $B$ out of all the transformations originating in $A$. The sum of the weights of all out-going edges at a given node must be 1. For example, suppose that the team transforms from the safe configuration cluster node $A$ to the $B, C,$ and $D$ safe configuration clusters. Suppose further that, historically, the team has shifted from configuration $A$ to configuration $B$ eight times, from $A$ to $C$ twelve times, and from $A$ to $D$ twenty times. Then $w(A, B)=\frac{8}{40}, w(A, C)=\frac{12}{40},$ and $w(A, D)=\frac{20}{40}$. \\

One may wish to remove all nodes of degree 0 (as they represent a cluster that is not part of any safe configuration transformation) and we do so in the examples below. Such clusters may have occurred when the team gained possession only briefly, made a long pass, or moved in a way not captured by our model.
\end{definition}

\begin{remark}[Markov chain interpretation]
With an assumption that configuration transformations do not depend on preceding transformations, $\mG_{\text{safe}}$ can be viewed as a Markov Chain where the safe configuration clusters represent the states and the weights represent the transition probabilities.
\end{remark}

We propose in \S \ref{sec:conclusion_and_further_questions} several uses for the safe progression graph.

\subsubsection{The adjacency matrix}\label{sss:adjacency_matrix}

 As the graph itself is rather unwieldy, we present $\mG_{\text{safe}}$ as an adjacency matrix where a row represents the start configuration node of a configuration transformation and the column is the end configuration node of the transformation. As such, there is a 1-to-1 map from the set of clusters to the set of rows (or columns) in the adjacency matrix of $\mG_{\text{safe}}$.
 Additionally, the clusters have been sorted according to the x-value of the cluster's center-of-mass, as defined in Definition \ref{def:cluster_com}. This sorting means that clusters which generally occur closer to the defensive net come before clusters which generally occur near the attacking net.

\begin{definition}[Cluster center-of-mass]\label{def:cluster_com}
Suppose that $C$ is a given cluster composed of $n$ passing matrices $\mathcal{M}=\{m_1, ..., m_n\}$. For each passing matrix, $m_i$, we denote the positions of the $l$ attacking players in the passing cliques of $m_i$ by $\{(x_{i, 1}, y_{i, 1}), ..., (x_{i, l}, y_{i, l})\}$. Then the center-of-mass of $C$, denoted $COM(C)$, is the pair $(\text{COM}_x, \text{COM}_y)$ defined by

\begin{equation}
    \text{COM}_x = \frac{1}{n}\sum_{i=1}^{n} \left( \frac{1}{l}\sum_{j=1}^{l}x_{i, j} \right)
    \hspace{0.5in} \text{and}
    \hspace{0.5in} \text{COM}_y = \frac{1}{n}\sum_{i=1}^{n} \left( \frac{1}{l}\sum_{j=1}^{l}y_{i, j} \right).
\end{equation}
\end{definition}

\begin{ex}[The safe progression graph adjacency matrix]

In this example we give the $\mG_{\text{safe}}$ adjacency matrix (Figure \ref{fig:Ex1SC_Graph}), computed using OPTICS clustering on the passing matrices for half the games of the data set of \S \ref{sec:data} in which Mjällby AIF played. More precisely, we used OPTICS in scikit-learn  \cite{OPTICS}. For clarity of the image, for the sake of Figure \ref{fig:Ex1SC_Graph} we iteratively removed all nodes of $\mG_{\text{safe}}$ of outgoing valence (not counting contributions to the valence by loops) $<2$.

\begin{figure}[ht!]
    \centering
    \includegraphics[width=0.8\textwidth]{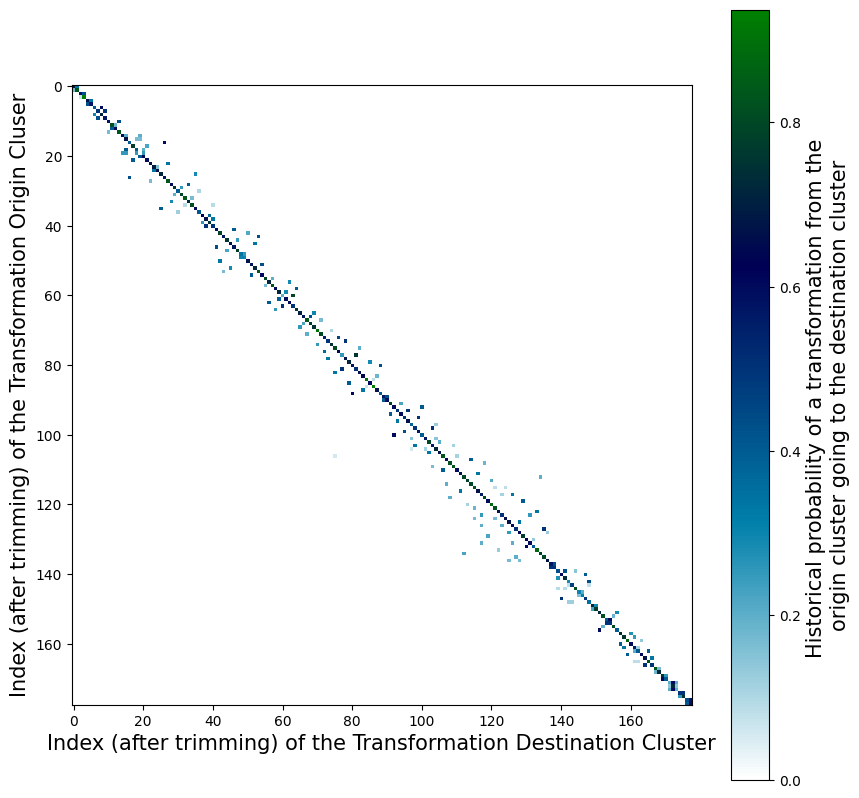}
    \caption{This figure displays the adjacency matrix of the safe progression graph obtained from half the Mjällby AIF games in the \S \ref{sec:data} data set. The axes represent the same set of clusters, which have been ordered by the mean $x$ value from the centers of mass of the passing cliques in each cluster. For example, passing cliques within the cluster referenced by $160$ on the two axes tend to be further up the field (toward the opposing end) compared to passing cliques in the $100$ cluster. A value of $v$ at entry $(x_i, y_j)$ indicates that $v \times 100\%$ of the transformations originating from cluster $y_j$ go to cluster $x_i$.
    }
    \label{fig:Ex1SC_Graph}
\end{figure}

We first observe that most of the transformations in Figure \ref{fig:Ex1SC_Graph} occur along, or near, the diagonal. Transformations directly on the diagonal correspond to transformations within the same cluster (i.e. the start and end configurations of the transformation are in the same cluster, so that the transformation forms a loop in $\mG_{\text{safe}}$). Transformations near the diagonal correspond to transformations between safe configurations that \emph{tend} to be located in the same general area of the field. It makes sense that such transformations make up the majority of the transformations in Figure \ref{fig:Ex1SC_Graph}, as most ground passes do not cover large distances. Transformations below the diagonal represent progression up the field.

\end{ex}

\FloatBarrier

\subsection{The Lose the Ball Node}\label{sec:LTB_Node_Main}

In the calculation of $\mG_{\text{safe}}$, we also recorded which clusters occurred directly before a loss of possession. This was encoded in $\mG_{\text{safe}}$ by adding a new node, called the \emph{Lose The Ball (LTB) node}, and then keeping track of the transformations during which the ball was lost via edges from a cluster node to the LTB node. Framing this into our Markov chain interpretation of $\mG_{\text{safe}}$, we have added a new state (LTB node) and adjusted the transition probabilities accordingly. The LTB node can be viewed as an absorbing state.

\begin{ex}[The LTB Node and the Mjällby AIF data]

If in $\mG_{\text{safe}}$ we only consider transitions to the LTB node or state, we are left with the transitions in Figure \ref{fig:Ex1LTB}. Figure \ref{subfig:Ex1LTB} shows the absolute number of transitions from each cluster to the LTB node and Figure \ref{subfig:Ex1LTB_norm} shows the transitions to the LTB node per visit to each state. In this example we did not require that nodes have an outgoing valence $\geq 2$.

\begin{figure}[ht!]
    \centering
    \begin{subfigure}[b]{1.0\textwidth}
        \centering
        \includegraphics[width=0.9\textwidth]{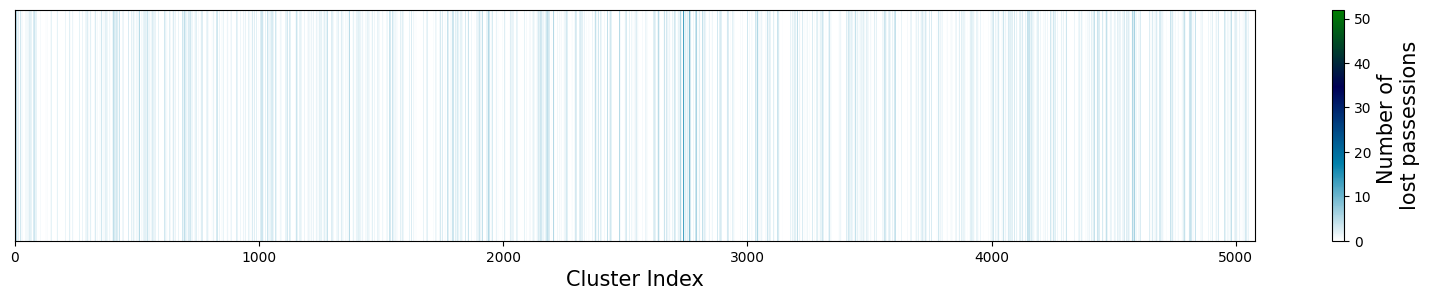}
        \caption{The adjacency matrix of all incoming edges to the Lose The Ball node. In this single-row matrix, a value of $v$ at cluster $x$ indicates that $v$ of the plays in cluster $x$ led to a loss of possession over the course of the entire season.}
        \label{subfig:Ex1LTB}
    \end{subfigure}
    \begin{subfigure}[b]{1.0\textwidth}
        \centering
        \includegraphics[width=0.9\textwidth]{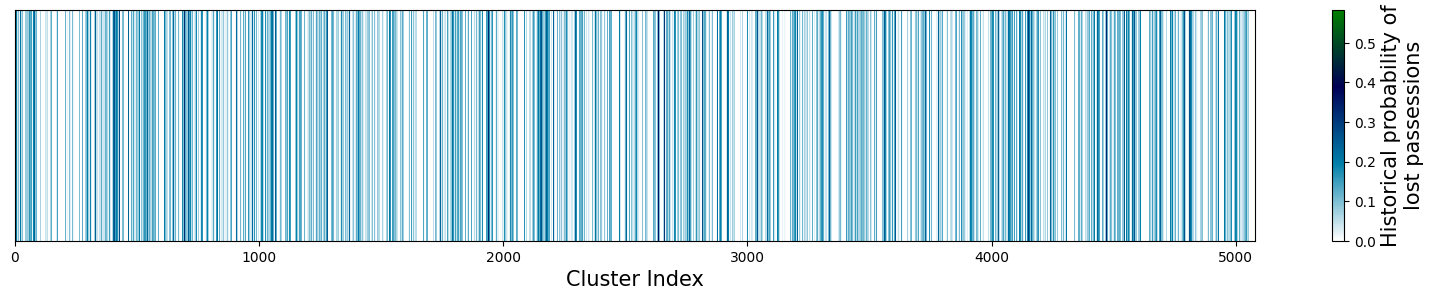}
        \caption{The adjacency matrix with the probability that a cluster transforms to the Lose The Ball node. A value of $v$ at cluster $x$ indicates that of all the plays in cluster $x$, $v \times 100\%$ of them resulted in a loss of possession.}
        \label{subfig:Ex1LTB_norm}
    \end{subfigure}
    \caption{This diagram records clusters occurring directly before a loss of possession, i.e. directly before the LTB node. One can identify likely ``problematic'' configurations occurring via darker lines.}
    \label{fig:Ex1LTB}
\end{figure}

 \FloatBarrier

We include in the following image 3 configurations that with high frequency lead into the LTB node.

\begin{figure}[ht!]
\centering
         \includegraphics[height=2in]{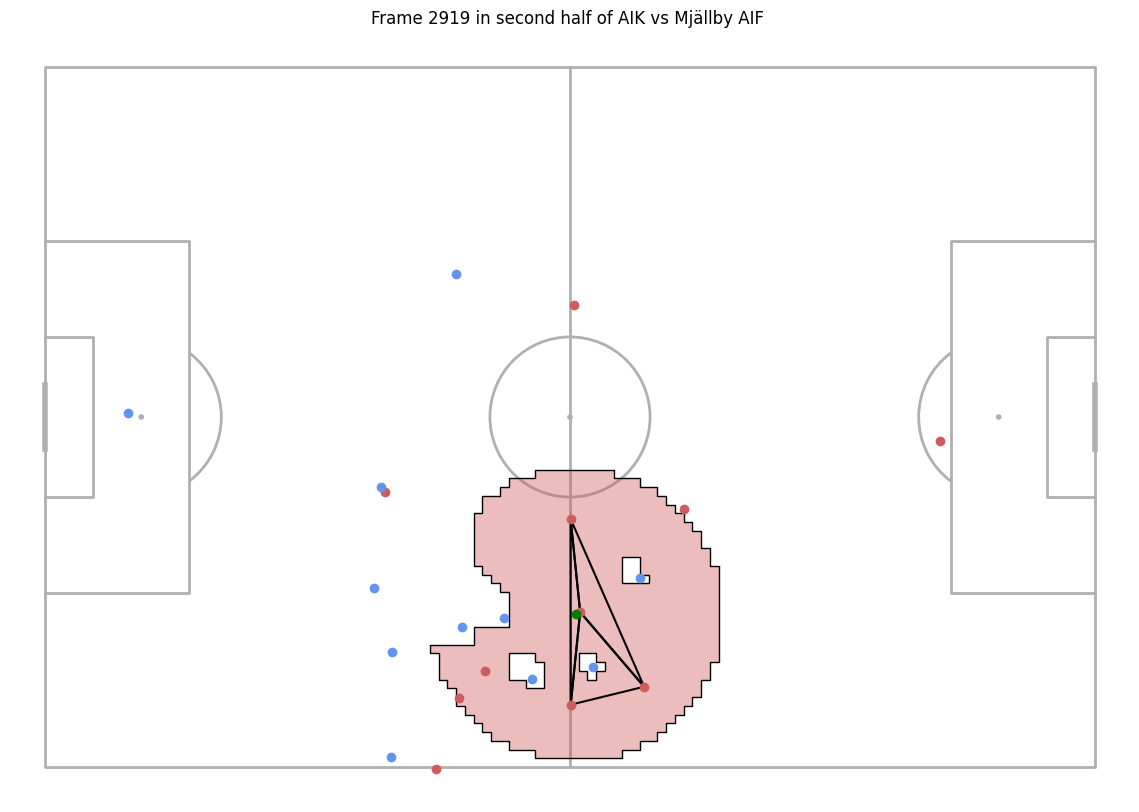}
         \includegraphics[height=2in]{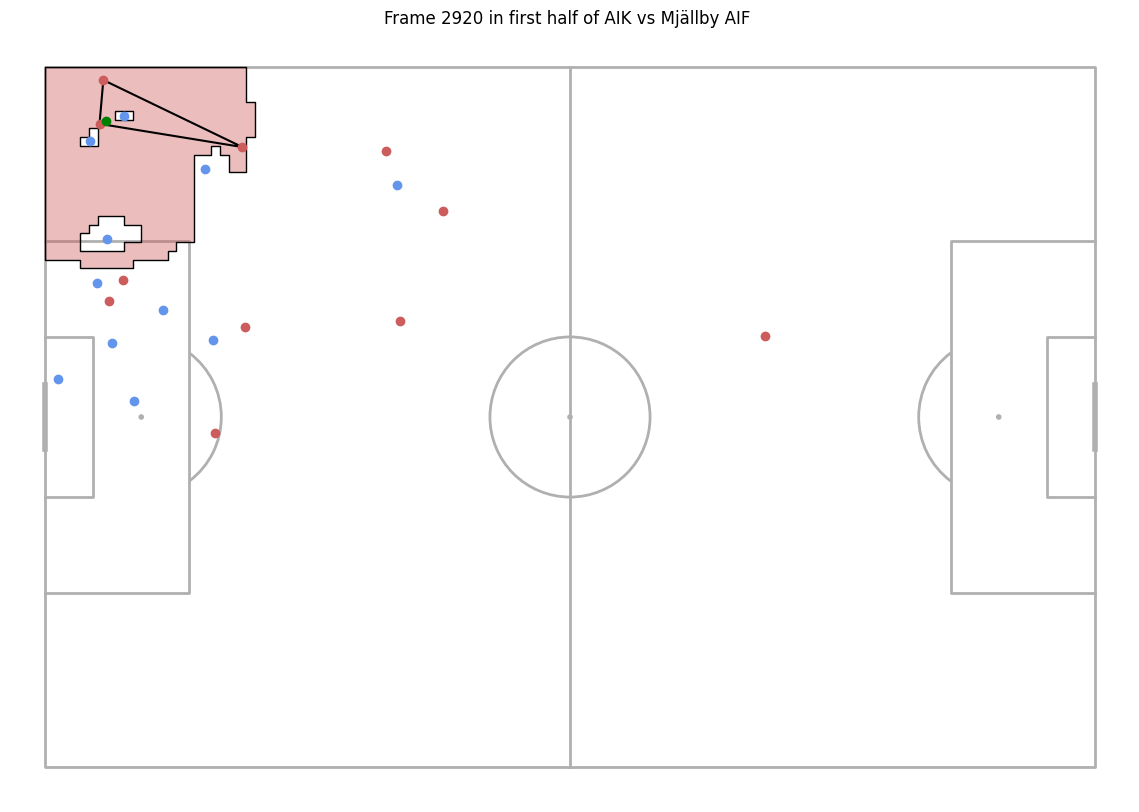}
 \end{figure}

 \begin{figure}[ht!]
\centering
         \includegraphics[height=2in]{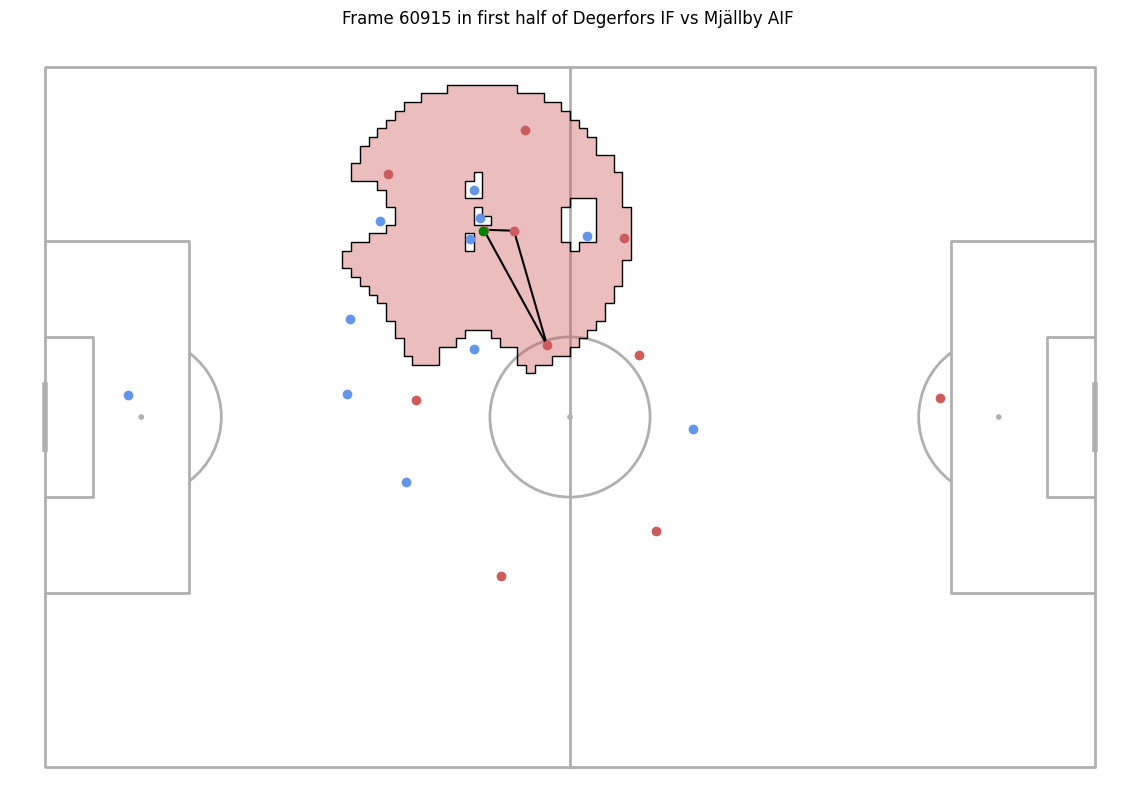}
 \end{figure}

 \FloatBarrier

 \end{ex}

\noindent The locations on the field are different, but one may notice that in each of these 3 frames:

1. there is only a single maximal passing clique and

2. the players in the clique without the ball give no progressive pass options to the player with the ball.

\noindent We did not run an analysis on whether (1) and (2) are typical for clusters directly preceding the LTB node, though one could particularly see how (2) could lead to an increased likelihood of losing possession, particularly if the players were eager to progress the ball and tried an option outside the clique.

 Notice that in each cluster the ball carrier is mostly surrounded by opposing blue players, and the radius of the passing cliques and passable region areas in these clusters are relatively small. Based off this it appears as though the defence is making a collectively aggressive movement, closing in on the player with the ball and moving up-field to cover opposing players who are in less dangerous positions, but are closer to the ball. This strategy is particularly visible in the last configuration pictured in. The blue team’s back line has moved up halfway to midfield despite the ball being on their side of the field. The right back in particular is very visibly leaving space between his defensive assignment on the weak wing, in order to provide added pressure on the ball. In addition, the blue team has three players closing in right on the ball, taking away any forward passing option. This strategy is aggressive and can lead to an opponent’s scoring chance if not timed properly, as it leaves the wide areas very vulnerable. However, we know this cluster is one of the most probable to reach the LTB node in the next progression, indicating that the blue team was mostly successful in pressing high on the ball.

\section{Clusters Preceding In-Box Possession}\label{sec:clusters_before_in_box_OPTICS}

\begin{figure}[ht!]
    \centering
    \includegraphics[width=0.9\textwidth]{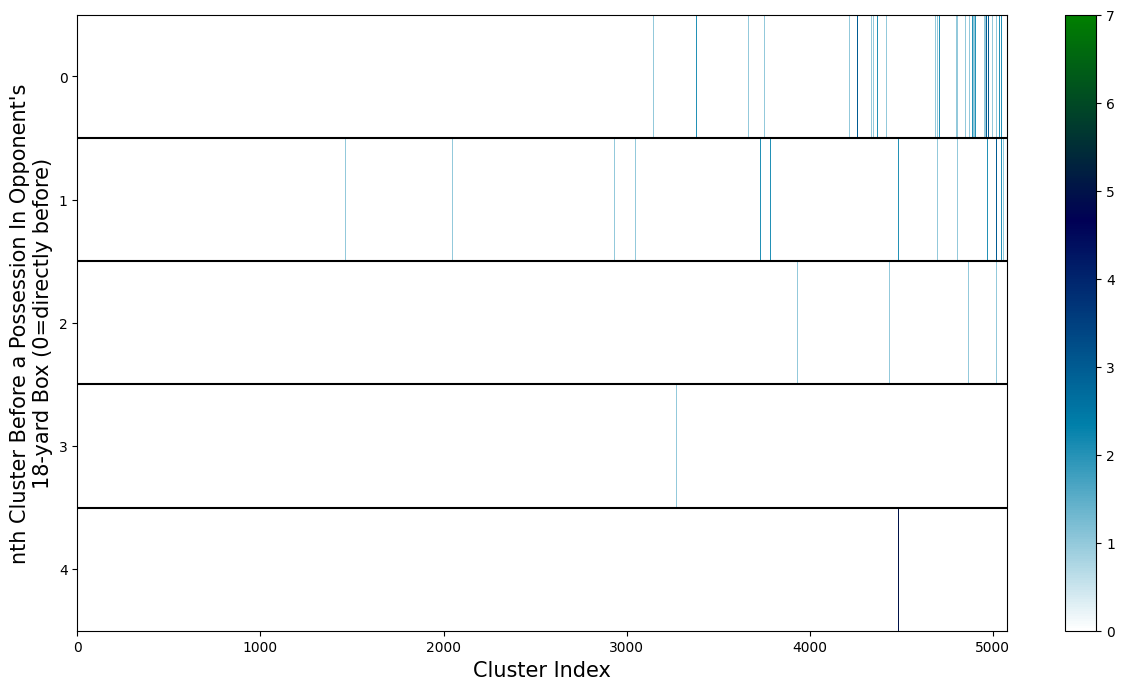}
    \caption{Visual representation of the clusters occurring leading up to a possession in the opponent's box. The $y$-axis corresponds to the temporal order of the cluster, where $y=0$ corresponds to a cluster occurring directly before a possession in the box. The $x$-axis corresponds to the same clusters as Figure
    \ref{fig:Ex1LTB}.}
    \label{fig:cluster_chains}
    \label{fig:OPTICS_chains}
\end{figure}

In contrast to focusing on the LTB node, one can observe the clusters that occur preceding a possession in the opponent's 18-yard box. A cluster was considered to occur directly before an in-box possession if it had a data point that occurred 1-10 seconds before the team entered the box with the ball (either a player dribbled into the box, or received a pass in the box). If multiple clusters had such data points, we considered the one that was closest in time to when they entered the box. That cluster was called \emph{Cluster 0}. From there, if a cluster had a datapoint that was 1-10 seconds \emph{before} Cluster 0, it was called Cluster 1. This continued, forming chains of clusters which occurred before the team had possession in the box. The clusters belonging to these chains are shown in Figure \ref{fig:cluster_chains}. The clusters are organized according to their mean center of mass, where clusters with larger indices correspond to clusters further up the field.

\FloatBarrier

\begin{ex}[Chain leading into the 18 yard box] This example includes a chain of safe configurations leading into the 18 yard box. Please see Figures \ref{fig:chain1} - \ref{fig:chain5}. As in Figure \ref{fig:OPTICS_chains}, we required that the time (in seconds) between each safe configuration was in $[1,10]$.

An interesting observation is that at all but one stage of the chain there is a player in the passing clique who is closer to the 18 yard box than the player with the ball. Even though that player is not always passed to, open passes not used can draw the attention of the opposition's defence, opening other options. A more thorough analysis can be found after the figures.

\begin{figure}[ht!]
    \centering
   \includegraphics[height=3.5in]{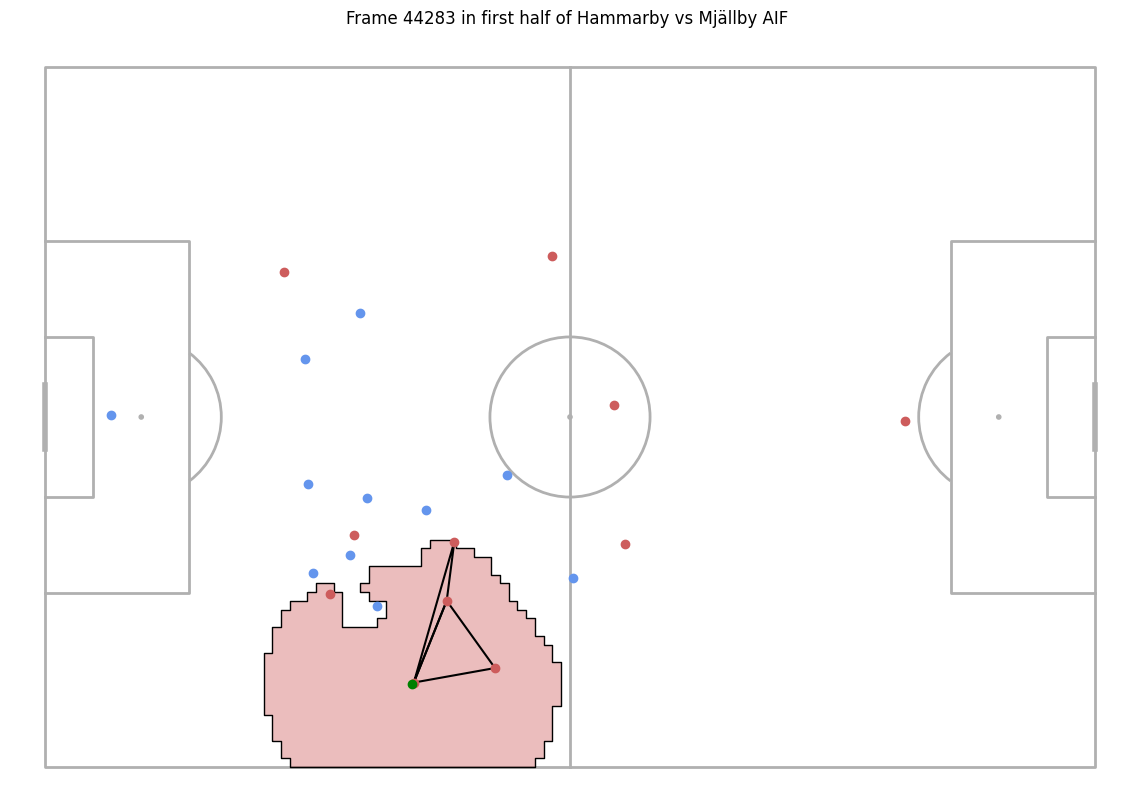}
   \caption{The first safe configuration in the chain }
   \label{fig:chain1}
\end{figure}

\begin{figure}[ht!]
    \centering
   \includegraphics[height=3.5in]{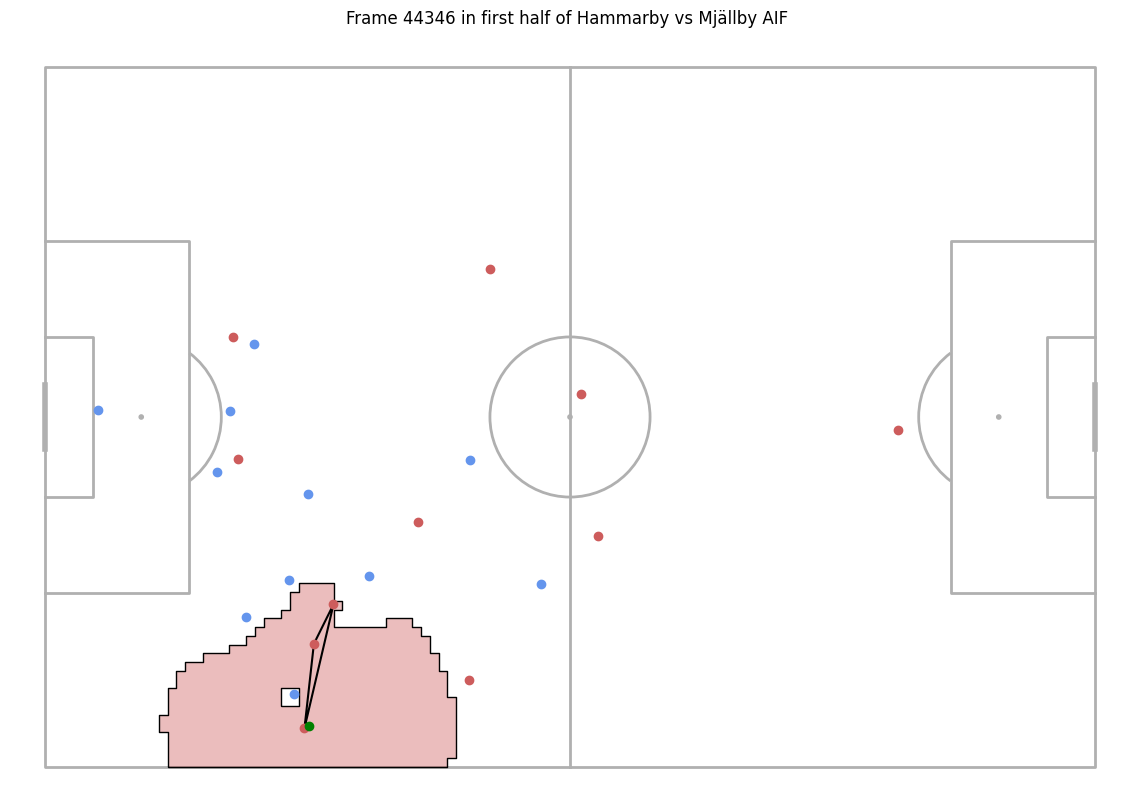}
   \caption{The second safe configuration in the chain}
   \label{fig:chain2}
\end{figure}

\begin{figure}[ht!]
    \centering
   \includegraphics[height=3.5in]{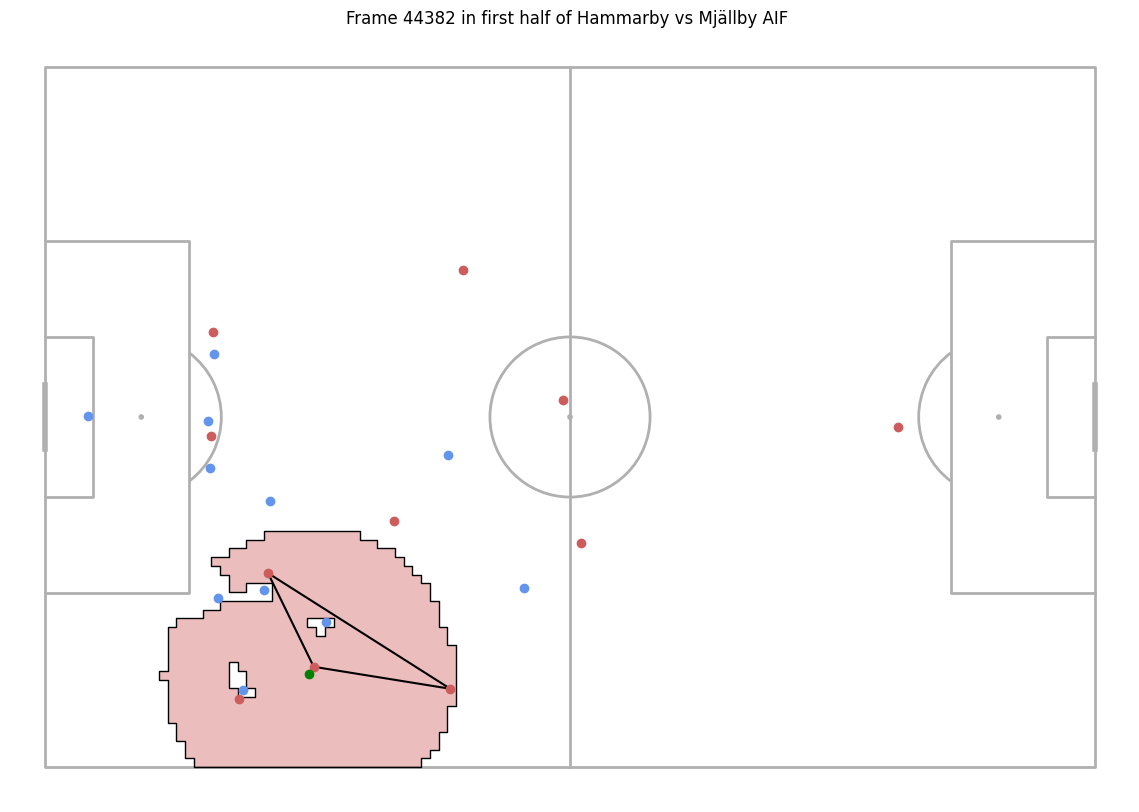}
   \caption{The third safe configuration in the chain}
   \label{fig:chain3}
\end{figure}

\begin{figure}[ht!]
    \centering
   \includegraphics[height=3.5in]{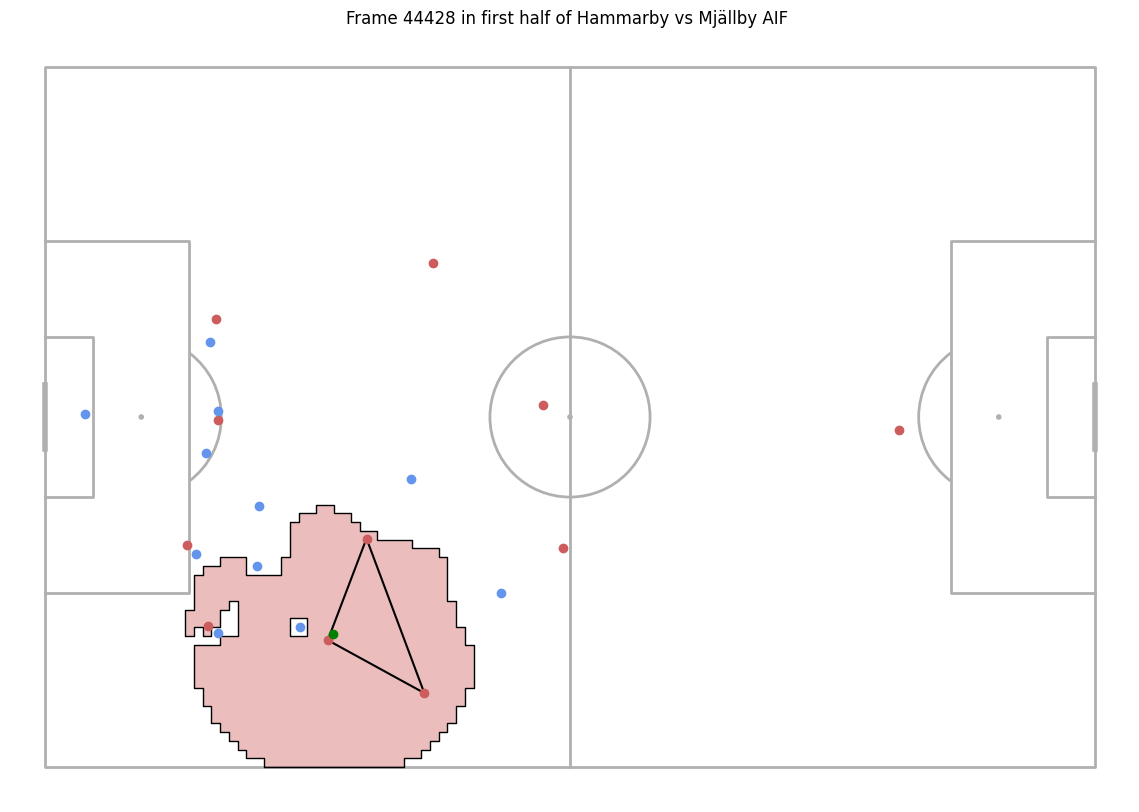}
   \caption{The fourth safe configuration in the chain (i.e. that that occurred directly before the possession in the box)}
   \label{fig:chain4}
\end{figure}

\newpage

\begin{figure}[ht!]
    \centering
   \includegraphics[height=3.5in]{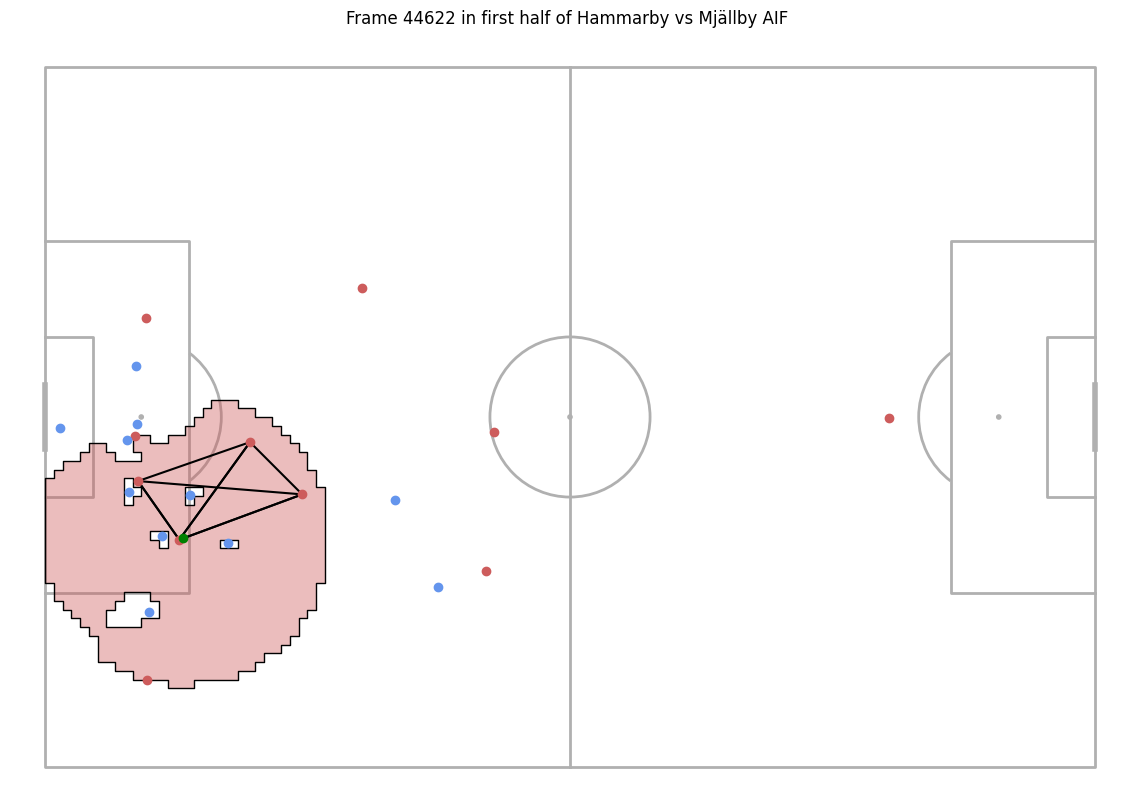}
    \caption{The fifth safe configuration in the chain (i.e. the first of the chain with possession in the box)}
   \label{fig:chain5}
\end{figure}

 \FloatBarrier

The possession chain begins in midfield with the ball carrier having 3 passing options (2 contained in the clique). Upon initial analysis it appears that the red team is completing a slow progression up the field, playing possession ball by bringing the majority of their midfield and attacking players close to the ball, opening up maximal passing options. This playing style is highlighted throughout the progression, as the attacking red team only uses short passes and quick movements to get possession in the 18-yard box. It is noteworthy that this offensive playing style requires the whole team to be very skilled at short accurate one touch passes and dribbling, as progressions usually involve most of the attacking players, and the ball must be moved around quickly since the defenders have shorter distances to cover.

During the configurations of Figures \ref{fig:chain1}-\ref{fig:chain2}, the ball carrier moves down the sideline, drawing pressure from the defender on that side. In the first configuration (Figure \ref{fig:chain1}), there are 6 blue defenders in the area of the passing clique. While this clique is a complete graph, any of the passing options receiving the ball are in a position to be immediately surrounded by 3 defenders. This clique is not stable (is likely not represented by a node with a loop in $\mG_{\text{safe}}$) and so the team needs to shift to another configuration, which they do: To dissipate this concentration of defenders, the furthest passing option of Figure \ref{fig:chain1} has moved away from the ball into the middle of the field, out of range of the passing clique, but successfully creating space for the other passing options in that area. This extra space proves beneficial, as the ball carrier passes to one of the open players from the configuration of Figure \ref{fig:chain2}, who now has space to run, look for a ball into the box, or continue to move it around the outside.

With the extra space the ball carrier waits and allows for other attacking players to make dangerous runs into the box. In Figure \ref{fig:chain4} a defender pressures the ball carrier leaving space behind him on the edge of the box exposed. The player who just passed off the ball takes advantage and runs into the space, receiving the ball just inside the box and with their momentum in the direction of the goal as shown in Figure \ref{fig:chain5}. This usually generates a high quality scoring opportunity, as the ball carrier is in range of the net, and has multiple passing options nearby. It can also be seen that the attacker who started behind the ball carrier in Figure \ref{fig:chain4}, also makes a run down the sideline into the space created by the player who received the ball in Figure \ref{fig:chain5}, giving the passer another dangerous option.

It is noteworthy that the constant movement of all players in this progression is what makes it so effective. When playing within a small radius of passing options, the opponent's defensive shape tightens as more passes are made, hence movement is required to keep breaking the defensive shape and cause miscommunication on defensive assignments.  This rapid progression through different configurations would be visible in $\mG_{\text{safe}}$ as a path without loops.

 \FloatBarrier

\end{ex}

\section{Future Possession Probability}\label{sec:future_possession_probability}

In this section we provide some evidence that safe player configurations, i.e. those with passing cliques, have a higher probability of maintaining possessions than player configurations without passing cliques.

For our test, we considered the following 3 sets of selection criteria for selecting a random frame:\\
\noindent {\bf{Criteria 1:}} possession is defined\footnote{From our observations, it appeared the possession was considered undefined when either the ball location was missing from the data or there was no clear ball possessor.} and the team in possession has a passing clique,\\
\noindent {\bf{Criteria 2:}} possession is defined and the team in possession does not have a passing clique, and\\
\noindent {\bf{Criteria 3:}}  possession is defined with no preference for existence of a passing clique.

For each set of selection criteria, we took 3000 samples of random frames satisfying the selection criteria. To select each sample, we first randomly selected a game, then randomly selected a frame from that game which satisfied the given criteria.

For each sample frame, we identified the team in possession as Team 1, and the opposition as Team 2. We then looked at a $t$-second window, starting T seconds in the future (where T ranges from 1-39 and $t\in \{1,2,3,4\}$), and logged it as a success/failure according to the following scheme: \\
\noindent  $\bullet$ If Team 1 had possession at least once in the t-second window, and Team 2 never had possession in that window, then the sample was logged as a success. \\
\noindent  $\bullet$ If the Team 2 had possession at least once during the t-second window, and Team 1 never did, then the sample was logged a failure. \\
\noindent  $\bullet$ If both teams had possession during the t-second window, or if possession was undefined for the entire window, the sample was rejected and not counted as one of the 3000 samples taken.

The columns in the following tables represent: \\
\noindent {\bf{time (s):}} the parameter T, in seconds, \\
\noindent {\bf{diff\%:}} (success rate for a random frame with a clique) - (success rate for a random frame without a clique) divided by (the success rate for a random frame without a clique)

The results of this test are displayed in Table 2 - Table 5. The main positive result is that ``diff\%'' is consistently positive, indicating that the existence of the clique improves one's chances of having possession in the future.

\FloatBarrier

\begin{table}[ht!]
\centering
\caption{Future Possession Probability, with 1-4 second windows (full tables in the appendix)}
\begin{tabular}{rrrrrrrr}
\toprule
 time T (s) &  t=1 diff \% &  t=2 diff \% & t=3 diff \% &  t=4 diff \% \\
\midrule
        1 &  -0.612 &  0.358 &  0.288 &  2.346 \\
        3 &  -1.693 &  1.297 & -0.395 &  0.197  \\
        5 &  -0.924 &  0.500 &  0.042 &  1.354 \\
        7 &  0.350 &  4.014 &  1.455 &  2.118 \\
        9 &  1.134 &  0.555 &  0.140 &  3.603 \\
       11 &  2.748 &  0.287 &  1.812 &  0.624 \\
       13 &  5.497 &  1.670 &  6.074 &  2.869 \\
       15 &  4.573 &  4.146 &  6.920 &  5.882 \\
       17 &  8.481 &  5.016 &  5.398 &  5.692 \\
       19 &  6.171 &  4.348 &  4.729 &  3.243 \\
       21 &  7.200 &  9.218 &  8.311 &  5.504 \\
       23 &  9.858 &  5.543 &  4.087 &  6.798 \\
       25 &  9.918 &  4.030 &  3.555 &  8.424 \\
       27 &  4.654 &  3.970 &  8.030 &  1.078 \\
       29 & 6.775 &  5.430 &  3.580 &  3.642 \\
       31 &  7.816 & -0.426 &  2.542 &  4.287 \\
       33 &  6.224 &  3.889 &  6.235 &  3.195 \\
       35 &  1.641 &  4.603 &  0.063 &  2.222 \\
       37 &  4.290 &  9.638 &  5.584 &  0.588 \\
       39 & -1.136 &  5.729 &  4.400 & -2.295 \\
\bottomrule
\end{tabular}
\end{table}

\FloatBarrier

One may be surprised that the difference between success rates with and without cliques stays positive even after 40 seconds. One possible explanation is that a team that had one clique is more likely to have another during those 40 seconds, leading to an increased likelihood of maintaining possession (based on the statistics for lower T values).

\section{Conclusions and Future Work}\label{sec:conclusion_and_further_questions}

\subsection{Data Limitations}\label{sec:data_limitations}

The position of the ball and the team in possession were missing from large sections of the data (in particular, approximately $36\%$ of the frames were missing the ball location). Hence, we could not determine the passing cliques in the moments of time represented by these frames. As higher quality tracking data becomes more available, one could (for example) better track the efficacy and ubiquity of the use of paths in the safe progression graphs, as well as get a more accurate computation of the graphs.

\begin{figure}[ht]
    \centering
    \includegraphics[width=7cm]{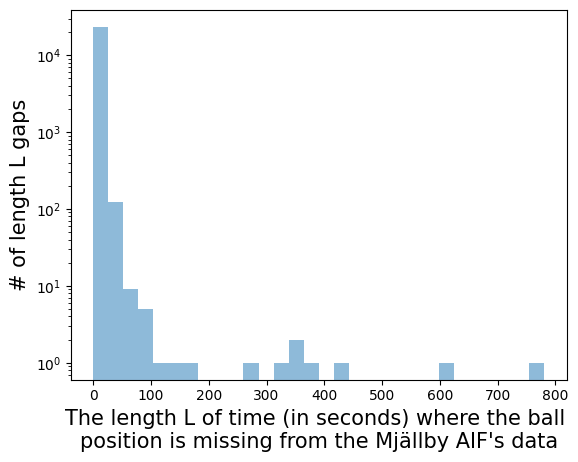}
    \caption{ }
    \label{fig:DataIsWonky}
\end{figure}

\FloatBarrier

\subsection{Future directions}\label{sec:future_directions}

We pose here several additional questions for further research. First, our framework does not consider opponent response. If we can develop an understanding of how a particular defending team is likely to respond to particular attacking moves, that could lead to a more useful decision-support tool. Second, building on the previous question, it would be useful to create valuations of different paths in the safe possession graph, where those valuations could be based on a suitable metric, such as expected goals, one based on a certain ``robustness index'' determined by the outgoing valence of each vertex in the path, by the EPV, or by some combination of these metrics. Such valuations could inform a dynamic decision making model of how best to traverse the safe possession graph from the  defending third to the attacking third. Third, one could determine player evaluation metrics based on the role they play in high value paths in the safe progression graph. Finally, more work needs to be done to demonstrate empirically the value of safe configurations, extending the initial results we have provided in this paper. For example, with the data we had access to, it was difficult to determine how frequently paths in the safe progression graph were successfully used (particularly by successful teams). Such analysis would still be valuable.

\subsection{Potential Coaching Applications}\label{sec:coaching_applications}

In this section we go over some potential uses by coaches of the work described in this manuscript.

\subsubsection{Applications of the safe progression graph and Player Configuration Metrics}\label{sec:applications}

One could also define the \emph{Attacking Configuration Robustness Index (ACRI)} for a node $A$ in a safe progression graph $G$, denoted $ACRI(A)$, as the sum of the weights of all out-going edges at $A$:
\begin{equation}
    ACRI(A) = \sum_{B\in G}^{}w(A,B)
\end{equation}
The inspiration behind the ACRI is as follows.
Given a vertex representing a configuration C of attacking players, an outgoing edge (from $C$ to $C'$) of high weight represents a likely transformation from $C$ to $C'$. Thus, a vertex $C$ having high weighted outgoing valence indicates that the corresponding configuration $C$ likely has multiple good options for transforming into other ``safe configurations.''

Suppose that for a particular team's safe progression graph, the weight on edge $E(X, LTB)$ is high. This could indicate that cluster $X$ is ``dangerous'' and should be avoided. This can be used to design training sessions to focus on these areas of weakness.

On the other hand, the defensive team can analyze the safe progression graph of their opponents to determine player configurations that they should either avoid allowing the opposition to enter or learn how to defensively manage. These nodes to look out for would be nodes $A$ that

\begin{enumerate}
    \item have a low $(A,LTB)$ weight, and/or
    \item have a large $ACRI(A)$, and/or
    \item lead to many goals in the sense of Figure \ref{fig:cluster_chains}.
\end{enumerate}

In particular, a large ACRI indicates multiple potential means of progressing out of the configuration, which it may not be possible to simultaneously defensively prevent or ``contain.'' One should be careful in this assumption, though, because a configuration whose node is of high ACRI may be easier to defensively manage than it may first appear as it is possible that multiple outgoing transformations could be defensively managed in the same manner.

With the previous paragraph in mind, one could also find paths in which each node has a high ACRI, making the sequence of safe configurations particularly difficult to defend against.

\subsubsection{Node Separating Edges}\label{sec:node_separating_edges}

The use of node separating edges, introduced in Definition \ref{def:node_separating_edge}, can be applied in interesting ways to the \Graph. We give one example of such an application below the definition.

\begin{definition}[Node separating edge]\label{def:node_separating_edge}
Given an edge set $E$, let $E-E'$ denote the set of edges in $E$ not in $E'$.
Suppose we have a safe progression graph $G=(V, E)$ and are given two nodes $X,Y\in G$. An edge \emph{separating $X$ and $Y$} in $G$ is an edge $E' \in E$ such that $G=(V, E-E')$ contains no path from $X$ to $Y$.
\end{definition}

Given a node in the defensive third and a node in the attacking third, an edge separating those two nodes could highlight which cluster transformations should be successfully defended against (i.e. the defence gains possession) in order to ensure the attacking team never reaches that attacking third configuration.

\section{Appendix: Tables}{\label{Appendix}}

We include here the tables from the computations described in \S \ref{sec:future_possession_probability}.

The columns in the following tables represent: \\
\noindent {\bf{time (s):}} the parameter T, in seconds, \\
\noindent {\bf{w/ clique:}} the success rate for a random frame with a clique \\
\noindent {\bf{w/o clique:}} the success rate for a random frame without a clique\\
\noindent {\bf{base:}} the success rate from a random frame with defined possession\\
\noindent {\bf{A B diff:}} A-B, where A and B are one of the previously mentioned columns. For example, ``w/ w/o diff'' is the difference between success rates of ``with cliques'' and ``without cliques''.\\
\noindent {\bf{diff\%:}} ``w/ w/o diff'' divided by ``w/o cliques''

\begin{table}[ht!]
\centering
\caption{Future Possession Probability, with 1 second window}
\begin{tabular}{rrrrrrrr}
\toprule
 time (s) &  w/ cliques &  w/o cliques &     base &  w/ w/o diff &  w/ base diff &  w/o base diff &    diff \% \\
\midrule
        1 &    0.926 &     0.932 & 0.920 &    -0.00567 &      0.00633 &       0.0120 & -0.612 \\
        3 &    0.847 &     0.861 & 0.850 &    -0.0143 &     -0.00300 &       0.0113 & -1.693 \\
        5 &    0.794 &     0.801 & 0.796 &    -0.00733 &     -0.00267 &       0.00467 & -0.924 \\
        7 &    0.762 &     0.760 & 0.755 &     0.00267 &      0.00767 &       0.00500 &  0.350 \\
        9 &    0.735 &     0.727 & 0.730 &     0.00833 &      0.00533 &      -0.00300 &  1.134 \\
       11 &    0.691 &     0.672 & 0.697 &     0.0190 &     -0.00533 &      -0.0243 &  2.748 \\
       13 &    0.691 &     0.653 & 0.666 &     0.0380 &      0.0257 &      -0.0123 &  5.497 \\
       15 &    0.663 &     0.633 & 0.649 &     0.0303 &      0.0147 &      -0.0157 &  4.573 \\
       17 &    0.656 &     0.601 & 0.615 &     0.0557 &      0.0413 &      -0.0143 &  8.481 \\
       19 &    0.632 &     0.593 & 0.602 &     0.0390 &      0.0297 &      -0.00933 &  6.171 \\
       21 &    0.625 &     0.580 & 0.582 &     0.0450 &      0.0430 &      -0.00200 &  7.200 \\
       23 &    0.609 &     0.549 & 0.580 &     0.0600 &      0.0290 &      -0.0310 &  9.858 \\
       25 &    0.608 &     0.548 & 0.554 &     0.0603 &      0.0543 &      -0.00600 &  9.918 \\
       27 &    0.573 &     0.546 & 0.556 &     0.0267 &      0.0167 &      -0.0100 &  4.654 \\
       29 &    0.576 &     0.537 & 0.534 &     0.0390 &      0.0363 &      -0.00267 &  6.775 \\
       31 &    0.559 &     0.515 & 0.544 &     0.0437 &      0.0147 &      -0.0290 &  7.816 \\
       33 &    0.562 &     0.527 & 0.519 &     0.0350 &      0.0430 &       0.00800 &  6.224 \\
       35 &    0.548 &     0.539 & 0.508 &     0.00900 &      0.0400 &       0.0310 &  1.641 \\
       37 &    0.528 &     0.506 & 0.518 &     0.0227 &      0.0100 &      -0.0127 &  4.290 \\
       39 &    0.528 &     0.534 & 0.503 &    -0.00600 &      0.0257 &       0.0317 & -1.136 \\
\bottomrule
\end{tabular}
\end{table}

\begin{table}[ht!]
\centering
\caption{Future Possession Probability, with 2 second window}
\begin{tabular}{rrrrrrrr}
\toprule
 time (s) &  w/ cliques &  w/o cliques &     base &  w/ w/o diff &  w/ base diff &  w/o base diff &    diff \% \\
\midrule
        1 &    0.930 &     0.927 & 0.937 &     0.00333 &     -0.00700 &      -0.0103 &  0.358 \\
        3 &    0.848 &     0.837 & 0.858 &     0.0110 &     -0.0100 &      -0.0210 &  1.297 \\
        5 &    0.801 &     0.797 & 0.802 &     0.00400 &     -0.00167 &      -0.00567 &  0.500 \\
        7 &    0.772 &     0.741 & 0.751 &     0.0310 &      0.0217 &      -0.00933 &  4.014 \\
        9 &    0.720 &     0.716 & 0.711 &     0.00400 &      0.00833 &       0.00433 &  0.555 \\
       11 &    0.697 &     0.695 & 0.689 &     0.00200 &      0.00867 &       0.00667 &  0.287 \\
       13 &    0.679 &     0.667 & 0.673 &     0.0113 &      0.00533 &      -0.00600 &  1.670 \\
       15 &    0.659 &     0.632 & 0.646 &     0.0273 &      0.0137 &      -0.0137 &  4.146 \\
       17 &    0.631 &     0.600 & 0.616 &     0.0317 &      0.0153 &      -0.0163 &  5.016 \\
       19 &    0.621 &     0.594 & 0.590 &     0.0270 &      0.0307 &       0.00367 &  4.348 \\
       21 &    0.618 &     0.561 & 0.572 &     0.057 &      0.0460 &      -0.0110 &  9.218 \\
       23 &    0.601 &     0.568 & 0.559 &     0.0333 &      0.0420 &       0.00867 &  5.543 \\
       25 &    0.579 &     0.556 & 0.563 &     0.0233 &      0.0160 &      -0.00733 &  4.030 \\
       27 &    0.571 &     0.548 & 0.552 &     0.0227 &      0.0193 &      -0.00333 &  3.970 \\
       29 &    0.571 &     0.540 & 0.540 &     0.0310 &      0.0313 &       0.000333 &  5.430 \\
       31 &    0.548 &     0.550 & 0.535 &    -0.00233 &      0.0130 &       0.0153 & -0.426 \\
       33 &    0.540 &     0.519 & 0.517 &     0.0210 &      0.0230 &       0.00200 &  3.889 \\
       35 &    0.529 &     0.504 & 0.514 &     0.0243 &      0.0143 &      -0.0100 &  4.603 \\
       37 &    0.543 &     0.491 & 0.508 &     0.0523 &      0.0347 &      -0.0177 &  9.638 \\
       39 &    0.524 &     0.494 & 0.495 &     0.0300 &      0.0290 &      -0.00100 &  5.729 \\
\bottomrule
\end{tabular}
\end{table}

\begin{table}[ht!]
\centering
\caption{Future Possession Probability, with 3 second window}
\begin{tabular}{rrrrrrrr}
\toprule
 time (s) &  w/ cliques &  w/o cliques &     base &  w/ w/o diff &  w/ base diff &  w/o base diff &    diff \% \\
\midrule
        1 &    0.926 &     0.923 & 0.916 &     0.00267 &      0.00933 &       0.00667 &  0.288 \\
        3 &    0.845 &     0.848 & 0.853 &    -0.00333 &     -0.00867 &      -0.00533 & -0.395 \\
        5 &    0.793 &     0.792 & 0.797 &     0.000333 &     -0.00467 &      -0.00500 &  0.042 \\
        7 &    0.756 &     0.745 & 0.744 &     0.0110 &      0.0123 &       0.00133 &  1.455 \\
        9 &    0.714 &     0.713 & 0.704 &     0.00100 &      0.0100 &       0.00900 &  0.140 \\
       11 &    0.699 &     0.686 & 0.692 &     0.0127 &      0.00700 &      -0.00567 &  1.812 \\
       13 &    0.675 &     0.634 & 0.663 &     0.0410 &      0.0120 &      -0.0290 &  6.074 \\
       15 &    0.674 &     0.628 & 0.632 &     0.0467 &      0.0420 &      -0.00467 &  6.920 \\
       17 &    0.636 &     0.602 & 0.607 &     0.0343 &      0.0293 &      -0.00500 &  5.398 \\
       19 &    0.620 &     0.591 & 0.573 &     0.0293 &      0.0470 &       0.0177 &  4.729 \\
       21 &    0.610 &     0.559 & 0.571 &     0.0507 &      0.0383 &      -0.0123 &  8.311 \\
       23 &    0.595 &     0.571 & 0.562 &     0.0243 &      0.0333 &       0.00900 &  4.087 \\
       25 &    0.581 &     0.561 & 0.544 &     0.0207 &      0.0377 &       0.0170 &  3.555 \\
       27 &    0.586 &     0.539 & 0.541 &     0.0470 &      0.0443 &      -0.00267 &  8.030 \\
       29 &    0.559 &     0.539 & 0.543 &     0.0200 &      0.0157 &      -0.00433 &  3.580 \\
       31 &    0.538 &     0.524 & 0.533 &     0.0137 &      0.00433 &      -0.00933 &  2.542 \\
       33 &    0.561 &     0.526 & 0.514 &     0.0350 &      0.0470 &       0.0120 &  6.235 \\
       35 &    0.526 &     0.526 & 0.534 &     0.000333 &     -0.00767 &      -0.00800 &  0.063 \\
       37 &    0.519 &     0.490 & 0.506 &     0.0290 &      0.0137 &      -0.0153 &  5.584 \\
       39 &    0.530 &     0.507 & 0.513 &     0.0233 &      0.0177 &      -0.00567 &  4.400 \\
\bottomrule
\end{tabular}
\end{table}

\begin{table}[ht!]
\centering
\caption{Future Possession Probability, with 4 second window}
\begin{tabular}{rrrrrrrr}
\toprule
 time (s) &  w/ cliques &  w/o cliques &     base &  w/ w/o diff &  w/ base diff &  w/o base diff &    diff \% \\
\midrule
        1 &    0.938 &     0.916 & 0.917 &     0.0220 &      0.0207 &      -0.00133 &  2.346 \\
        3 &    0.846 &     0.844 & 0.834 &     0.00167 &      0.0117 &       0.0100 &  0.197 \\
        5 &    0.788 &     0.777 & 0.788 &     0.0107 &     -0.000333 &      -0.0110 &  1.354 \\
        7 &    0.755 &     0.739 & 0.736 &     0.0160 &      0.0197 &       0.00367 &  2.118 \\
        9 &    0.722 &     0.696 & 0.711 &     0.0260 &      0.0110 &      -0.0150 &  3.603 \\
       11 &    0.694 &     0.690 & 0.676 &     0.00433 &      0.0177 &       0.0133 &  0.624 \\
       13 &    0.662 &     0.643 & 0.652 &     0.0190 &      0.0107 &      -0.00833 &  2.869 \\
       15 &    0.652 &     0.613 & 0.618 &     0.0383 &      0.0333 &      -0.00500 &  5.882 \\
       17 &    0.638 &     0.602 & 0.608 &     0.0363 &      0.0303 &      -0.00600 &  5.692 \\
       19 &    0.617 &     0.597 & 0.587 &     0.0200 &      0.0297 &       0.00967 &  3.243 \\
       21 &    0.612 &     0.578 & 0.568 &     0.0337 &      0.0440 &       0.0103 &  5.504 \\
       23 &    0.593 &     0.553 & 0.585 &     0.0403 &      0.00833 &      -0.0320 &  6.798 \\
       25 &    0.582 &     0.533 & 0.548 &     0.0490 &      0.0333 &      -0.0157 &  8.424 \\
       27 &    0.556 &     0.550 & 0.542 &     0.00600 &      0.0143 &       0.00833 &  1.078 \\
       29 &    0.558 &     0.538 & 0.554 &     0.0203 &      0.00400 &      -0.0163 &  3.642 \\
       31 &    0.544 &     0.521 & 0.539 &     0.0233 &      0.00533 &      -0.0180 &  4.287 \\
       33 &    0.532 &     0.515 & 0.504 &     0.0170 &      0.0277 &       0.0107 &  3.195 \\
       35 &    0.525 &     0.513 & 0.513 &     0.0117 &      0.0123 &       0.000667 &  2.222 \\
       37 &    0.510 &     0.507 & 0.529 &     0.00300 &     -0.0193 &      -0.0223 &  0.588 \\
       39 &    0.508 &     0.520 & 0.513 &    -0.0117 &     -0.00433 &       0.00733 & -2.295 \\
\bottomrule
\end{tabular}
\end{table}

\FloatBarrier

\printbibliography 

\end{document}